%% file: main.tex
\documentclass{article}

\usepackage{PRIMEarxiv}
\usepackage{graphicx}      
\usepackage{amsmath,amssymb,amsfonts,amsthm}
\usepackage{booktabs}      
\usepackage{array}         
\usepackage{tabularx}      
\usepackage{multirow}     
\usepackage{wrapfig}
\usepackage{fancyhdr}      
\usepackage{lastpage}      
\usepackage{extramarks}    
\usepackage[title]{appendix}
\usepackage{xcolor}        
\usepackage{url}           
\usepackage{hyperref}      
\usepackage{listings}      
\usepackage{algorithm}     
\usepackage{algpseudocode} 
\usepackage{pifont} 
\usepackage{textcomp} 
\usepackage{tabularray}    
\usepackage{caption}       
\usepackage{manyfoot}      
\usepackage{orcidlink}
\usepackage{enumitem} 
\usepackage{fontawesome}
\usepackage{ragged2e} 
\usepackage{makecell} 

\usepackage{amsmath,amssymb}
\newcommand*\circled[1]{\textcircled{\raisebox{-0.9pt}{#1}}}

\definecolor{orange_bis}{RGB}{255,165,0}
\definecolor{blue_bis}{RGB}{0,0,255}
\definecolor{green_bis}{RGB}{0,128,0}
\definecolor{red_bis}{RGB}{255,0,0}

\usepackage[verbose=true,letterpaper]{geometry} 
\usepackage[numbers]{natbib}  

\usepackage[listings,breakable]{tcolorbox}
\usepackage{listings}

\definecolor{usercolor}{rgb}{1.0, 0.9, 0.9} 
\definecolor{assistantcolor}{rgb}{1.0, 1.0, 0.9} 
\definecolor{execcolor}{rgb}{0.8, 1.0, 0.77} 
\definecolor{toolcolor}{rgb}{0.95, 0.95, 0.95} 
\definecolor{codebg}{rgb}{0.1, 0.1, 0.1} 
\definecolor{codetext}{rgb}{0.8, 0.8, 0.8} 
\definecolor{white}{rgb}{1.0, 1.0, 1.0} 

\newtcolorbox{userbox}{%
  breakable,
  use color stack,
  colback=usercolor,
  colframe=magenta!50!black,
  boxrule=0.5mm,
  arc=3mm,
  left=2mm,
  right=2mm,
  top=1mm,
  bottom=1mm,
  width=\textwidth,
  sharp corners,
}

\newtcolorbox{assistantbox}{%
  breakable,
  use color stack,
  colback=assistantcolor,
  colframe=yellow!70!black,
  boxrule=0.5mm,
  arc=3mm,
  left=2mm,
  right=2mm,
  top=1mm,
  bottom=1mm,
  width=\textwidth,
  sharp corners,
}

\newtcolorbox{execbox}{%
  breakable,
  use color stack,
  colback=execcolor,
  colframe=green!70!black,
  boxrule=0.5mm,
  arc=3mm,
  left=2mm,
  right=2mm,
  top=1mm,
  bottom=1mm,
  width=\textwidth,
  sharp corners,
}

\newtcolorbox{toolbox}{%
  breakable,
  use color stack,
  colback=toolcolor,
  colframe=gray!70!black,
  boxrule=0.5mm,
  arc=3mm,
  left=2mm,
  right=2mm,
  top=1mm,
  bottom=1mm,
  width=\textwidth,
  sharp corners,
}

\lstdefinestyle{codeoutput}{
    backgroundcolor=\color{codebg},
    basicstyle=\ttfamily\footnotesize\color{white},
    keywordstyle=\color{white},
    stringstyle=\color{white},
    commentstyle=\color{white},
    breaklines=true,
    frame=none,
    rulecolor=\color{codebg}, 
    showstringspaces=false,
}

\newcommand{\cmark}{\textcolor{black}{\ding{51}}} 
\newcommand{\xmark}{\textcolor{red}{\ding{55}}}   

\title{\textbf{CliMB: An AI-enabled Partner for Clinical Predictive Modeling}}
\author{
  Evgeny Saveliev\thanks{These authors contributed equally.}~ \orcidlink{0000-0003-2887-0342}\\
  University of Cambridge \\
   \And
  Tim Schubert\footnotemark[1]~ \orcidlink{0009-0003-1696-4402} \\
  University of Cambridge \\
  University Hospital Heidelberg \\
  \And
  Thomas Pouplin \\
  University of Cambridge \\
  \vspace{10pt}
  \And
  Vasilis Kosmoliaptsis \orcidlink{0000-0001-7298-1387} \\
  University of Cambridge \\
  NIHR BTRU, Cambridge \\
   \And
  Mihaela van der Schaar \orcidlink{0000-0003-3933-6049} \\
  University of Cambridge \\
  \texttt{mv472@damtp.cam.ac.uk} \\
}
\date{April 2024}

\begin{document}
\maketitle

\newcommand\baseline{baseline GPT-4}
\newcommand\Baseline{Baseline GPT-4}

\begin{abstract}
    Despite its significant promise and continuous technical advances, real-world applications of artificial intelligence (AI) remain limited. We attribute this to the ``domain expert-AI-conundrum'': while domain experts, such as clinician scientists, should be able to build predictive models such as risk scores, they face substantial barriers in accessing state-of-the-art (SOTA) tools. While automated machine learning (AutoML) has been proposed as a partner in clinical predictive modeling, many additional requirements need to be fulfilled to make machine learning accessible for clinician scientists. 
    
    To address this gap, we introduce \textit{CliMB}, a no-code AI-enabled partner designed to empower clinician scientists to create predictive models using natural language. \textit{CliMB} guides clinician scientists through the entire medical data science pipeline, thus empowering them to create predictive models from real-world data in just one conversation. \textit{CliMB} also creates structured reports and interpretable visuals. In evaluations involving clinician scientists and systematic comparisons against a  \baseline, \textit{CliMB} consistently demonstrated superior performance in key areas such as planning, error prevention, code execution, and model performance. Moreover, in blinded assessments involving 45 clinicians from diverse specialties and career stages, more than 80\% preferred \textit{CliMB} over \baseline. Overall, by providing a no-code interface with clear guidance and access to SOTA methods in the fields of data-centric AI, AutoML, and interpretable ML, \textit{CliMB} empowers clinician scientists to build robust predictive models.

    The proof-of-concept version of \textit{CliMB} is available as open-source software on GitHub: \href{https://github.com/vanderschaarlab/climb}{https://github.com/vanderschaarlab/climb}.

\end{abstract}

\section{Introduction}
Individualized risk prediction is a key enabler of precision medicine \cite{Callender2023-ma}. There are many use cases for predictive models; we will highlight three illustrative examples. First, individualized survival predictions can guide the use of adjuvant therapy for patients with early invasive breast cancer \cite{Alaa2021-ur}. Second, predictive models can guide clinical pathways such as identifying high-risk patients during radiotherapy. Providing these individuals with supplemental clinical evaluations reduces costs while improving patient outcomes \cite{Natesan2024-gl}. Finally, risk predictions can also facilitate timely interventions, e.g. to reduce mortality in patients at risk of sepsis \cite{Kamran_Fahad2024-br}.

Machine learning (ML) methods offer increasingly powerful solutions for risk prediction. These include methods and tools to improve data quality \cite{Seedat2024-uw,Pleiss2020-eo,Yoon2018-yl,Jarrett2022-qt,Swayamdipta2020-ou,Smith2014-fb,Seedat2022-al}, optimize model performance \cite{Feurer2022-lw,Imrie2023-wp,Erickson2020-tp,Kotthoff2017-ky,Lindauer2021-xd}, and enhance transparency and interpretability \cite{Ghorbani2019-tf,Ribeiro2016-fm,Ghorbani2019-bb}. However with few exceptions, current clinical risk scores are either created with classical statistical methods or a small range of ML methods (random forests appear to dominate \cite{Wang2020-qv,Goldstein2017-oz}). This narrow focus is a problem in prediction because, as highlighted by the ``no free lunch'' theorem, no single method is deemed ideal for all prediction problems \cite{Wolpert1997-sv,Callender2023-ma}. Overall, the breadth of ML methodology being used in clinical risk prediction (and other applications) remains limited and thus the potential of AI in medicine unfulfilled \cite{Rajpurkar2022-zq}.

We argue that this is the result of a fundamental conundrum: on the one hand, domain experts (e.g., clinician scientists) should be able to build predictive models; on the other hand, there are largely impenetrable barriers to using SOTA methods. Solving this domain expert-AI-conundrum promises to be transformative for the medical field thanks to more individualized healthcare. Moreover, widespread application of predictive modeling in medicine will scale the identification of real-world needs that should be addressed by the ML community. Overall, a solution to the conundrum would enable progress in reality-centric AI adoption and development.

\subsection{Clinician scientists should be able to build predictive models}

\textbf{Why is clinical expertise crucial?} The primary goal of applying ML in the medical domain—and indeed in any domain—is not to explore difficult challenges or develop novel methods that achieve better performance on toy benchmarks. The true value of ML in healthcare lies in its potential to address real-world medical challenges and thereby improve patient outcomes. To ensure a predictive model will serve an actual need, the process of building it should begin with a clinical problem that requires predictions \cite{Balagopalan2024-lx}.
Clinicians are at the forefront of patient care with a deep understanding of medicine and awareness of clinical needs. They can identify a clear problem and setting where predictions would be promising \cite{Wynants2020-od}. Thus, in the clinical domain, clinicians should set the agenda for the development of predictive tools. The necessity of domain expertise is best illustrated by reflecting on the surge in ML models during the COVID-19 pandemic. At the time, hundreds of models with little to no impact were created, in part because efforts were not domain expert-driven \cite{Heaven2021-ll}.
Domain expertise is required not only to set the agenda but also throughout the development process. In addition, domain expertise is necessary for data exploration and engineering, which are usually the most time-intensive phases of data science projects \cite{Pfisterer2019-ub}.  In fact, many of the models created during the pandemic were built by AI researchers who lacked the medical expertise needed to identify flaws in the medical data \cite{Heaven2021-ll}. Although the subsequent model-building process may be more domain-agnostic, interpreting the validity of model predictions (e.g., based on feature importance) requires a deep understanding of the included clinical variables and measurements.

\textbf{Could you simply add a data scientist?} While creating dual-expert teams of clinician scientists and data scientists could be a potential solution to the domain expert-AI-conundrum, these teams are often costly, difficult to assemble, and prone to inefficiencies due to translation errors and misaligned goals \cite{Balagopalan2024-lx,Callender2023-ma,Mao2019-wz,Hou2017-rh}. In particular, errors in translation between ML researchers and clinician scientists can lead to problems like feature leakage, which result in impaired model quality. In addition, given that demographics and available features change over time, risk scores need to be updated on a regular basis. The necessity to repeat the predictive modeling cycle aggravates the problem, particularly considering the shortage of ML experts \cite{Van-der-Blom2021-zd}.

\begin{figure}[htbp]
    \centering
    \begin{minipage}{0.6\textwidth}
        \centering
        \includegraphics[width=\linewidth]{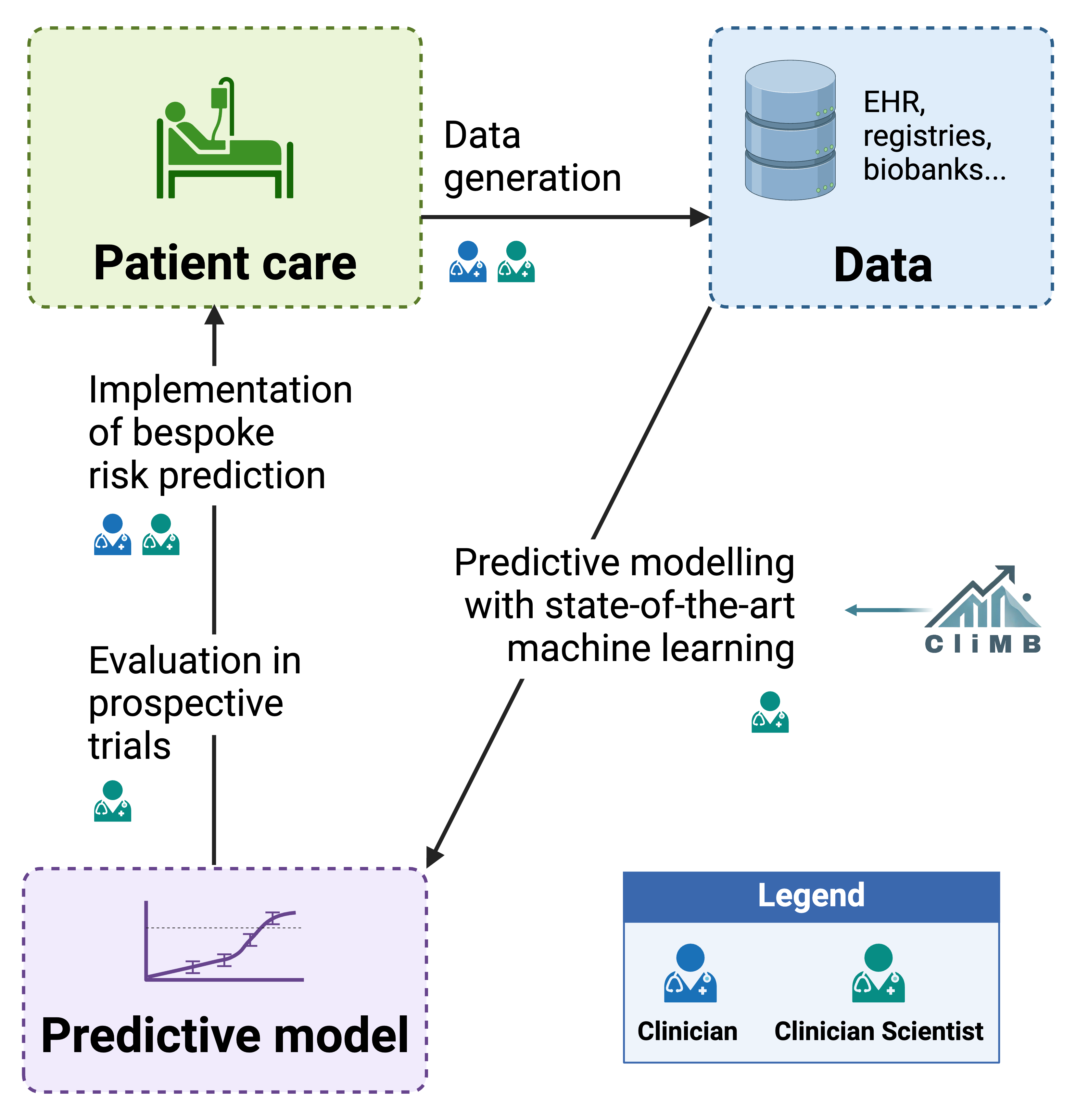}
    \end{minipage}%
    \hspace{0.02\textwidth} 
    \begin{minipage}{0.35\textwidth}
        \caption{\textbf{Cycle of clinical predictive modeling}. In this proposed cycle, data is generated during patient care by clinicians or clinician scientists. This data can be deposited in EHRs, registries, biobanks etc. Clinician scientists then create predictive models using SOTA machine learning. Empowering them to do so is the focus of this paper. Subsequently predictive models are evaluated and then implemented in patient care. This cycle will likely be repeated to adapt to changing features/demographics or new challenges. \vspace{0.5em} \\ \scriptsize{Figure created with Biorender.com}}
        \label{Cycle of predictive modeling}
    \end{minipage}
\end{figure}

\textbf{How would clinical risk scores be built in the future?} We envision a cycle in which clinicians or clinician scientists care for their patients, generate data in the process, clinician scientists use this data to build predictive models, evaluate these models in prospective trials, and implement them directly in patient care (see Figure \ref{Cycle of predictive modeling}). Putting domain experts in the driver’s seat is expected to (1) improve the clinical utility of ML tools, and (2) bridge the gap between method development and real-world application \cite{Balagopalan2024-lx}.

\subsection{Clinician scientists face largely impenetrable access barriers}

\textbf{Why is risk prediction hardly accessible?} Substantial resources and expertise are currently required to develop predictive models \cite{Callender2023-ma}. As explained above, no single ML method will be ideal for the variety of risk scores being built for healthcare. Expecting even a few clinician scientists engaged in routine patient care to keep up with the latest developments in ML and make informed choices about the most promising strategies for their prediction problem is unrealistic. 

\textbf{What has been done so far to make ML more accessible and why is it insufficient?} Substantial efforts have been made to ``democratize'' ML for domain experts with little to no ML knowledge, particularly in the field of automated machine learning (AutoML) \cite{Kotthoff2017-ky,Feurer2015-pz,Lindauer2022-ex}. While the AutoML community has created powerful tools to optimize the model building process and make it more systematic and efficient \cite{Hutter2019-ou}, AutoML does not deliver on its original purpose of democratizing ML. Over the past decade, AutoML achievements focused on optimizing predictive performance \cite{Lindauer2024-th}. However, current tools are not fully aligned with the needs of domain experts such as clinician scientists who seek to apply off-the-shelf ML solutions for their problems \cite{Lindauer2022-ex}. In general, user interaction with AutoML systems remains an underexplored area. In particular, the lack of sufficiently intuitive methods for users to formulate tasks, contribute domain expertise, and communicate preferences and constraints is a major obstacle for people with non-technical backgrounds \cite{Lindauer2024-th}. For instance, most AutoML tools today require technical understanding and skills (e.g., the ability to use Python packages like AutoPrognosis \cite{Imrie2023-wp}). As a result, these tools tend to (perhaps inadvertently) cater to ML experts seeking to enhance their workflow, rather than serving domain experts with limited AI knowledge \cite{Tornede2023-jg}. It has thus been proposed that natural language interfaces are a prerequisite for domain experts to interact comfortably with AutoML and other data science tools \cite{Karmaker_Santu2021-yv}.

\textbf{Which challenges exist beyond intuitive AutoML interfaces?} Usability is further limited because the tools required to build robust risk prediction models are fragmented rather than integrated into a comprehensive pipeline that clinician scientists could default to. The more comprehensive pipelines that do exist are focused on the model-building process but omit the critical and time-consuming phases of data exploration and engineering \cite{Tornede2023-jg,Kumar2024-dw}. Data quality largely determines the performance, fairness, robustness, safety, and scalability of AI systems. As a result, poor data quality can have disproportionately negative effects, particularly in high-stakes domains like medicine. Unfortunately, medical data is usually not ML-ready, making the data engineering phase critical for medical data science projects \cite{Sambasivan2021-ia,Balagopalan2024-lx}. Human-centric AI must also consider interpretability regarding the data, modeling process, and outcomes. Interpretability and transparency are crucial sources of trust in AutoML systems \cite{Lindauer2024-th,Wang2019-fu,Drozdal2020-hr}. While advanced interpretability tools have been developed \cite{Agarwal2020-ta}, they require technical skills to be used and knowledge to be understood. To truly benefit domain experts like clinician scientists, interpretability methods must be accompanied by contextual, easy-to-comprehend explanations \cite{Tornede2023-jg,Moosbauer2021-um}.

\subsection{Requirements for human-centric AI/ML partners for predictive modeling}\label{requirements}

To resolve the domain expert-AI-conundrum and make ML accessible to clinician scientists without the need for a data scientist, an AI-enabled partner is needed that builds on developments in AutoML, data-centric AI, and ML interpretability. We refer to this kind of tool as a partner rather than a ``co-pilot'' because it would not merely enhance a clinician scientist's (the pilot's) capabilities but rather augment the clinician scientist's skill set with a new class of capabilities. We consulted with clinician scientists, ML researchers, statisticians, and software engineers to distill the requirements that an ideal partner for predictive modeling in medicine would fulfill. They are listed below:

\begin{itemize}
    \item[\ding{172}] \textbf{Holistic support.} The partner should accompany the domain expert through all stages of their data science project. Before starting the project, an ideal partner would check the alignment between the user's goal, their data, hardware and software. For easy access and an (albeit modular) off-the-shelf solution, tools for all phases of the predictive modeling pipeline would be combined. These phases include (1) data exploration, (2) data engineering, (3) model building, and (4) model exploitation \cite{De_Bie2022-eu}. Holistic support encompasses not only execution but also planning, which is an underexplored area \cite{Gu2024-th}.
    \item[\ding{173}] \textbf{Usability.} To truly empower domain experts, the AI/ML partner must be user-friendly. Usability hinges on automation and interfaces. Most basic tasks should be automated, requesting human feedback only when critical decisions must be made. Communication should occur through natural language, a universal interface \cite{Zhang2023-of}. Additionally, the visual presentation must be clear and easy to navigate. Improved usability will also enhance interpretability and transparency.
    \item[\ding{174}] \textbf{Best practices.} Data science best practices should be followed to create predictive models that perform well, generalize effectively, and maintain clinical validity. E.g., model building should be optimized using AutoML. User misconceptions should be identified and interactively addressed. 
    \item[\ding{175}] \textbf{Robustness.} As evidenced in section \ref{exp_comparison_against_gpt4V}, large language models (LLMs) often deviate from their intended plans, leading to incomplete or suboptimal outcomes. A truly effective AI/ML partner, however, must adhere to robust yet adaptable strategies to prevent it from veering off course and neglecting critical steps.
    \item[\ding{176}] \textbf{Interpretability.} Interpretability and transparency should be integrated at all stages to allow the domain expert to critically appraise processes and outcomes. Interpretability metrics should be made accessible through interactive explanations, contextualization, and visualizations.
    \item[\ding{177}] \textbf{Modularity.} An ideal AI/ML partner should be designed to scale with future methodologies, modularly incorporate novel tools as they emerge, and accommodate existing tools the user wishes to integrate alongside off-the-shelf solutions.
    \item[\ding{178}] \textbf{Versatility.} Within reasonable boundaries, an AI/ML partner should dynamically adapt to user requests. For example, it should be possible to revisit previous phases of the data science pipeline if the need arises at a later stage of the project.
    \item[\ding{179}] \textbf{Privacy.} Privacy is a key concern in medicine \cite{Price2019-zc}. If sensitive medical data are used for data science tasks, appropriate privacy features must be an integral part of any tools that access these data.
\end{itemize}

\subsection{An AI-enabled partner for clinical predictive modeling}
\begin{figure}[htbp]
    \centering
    \begin{minipage}{0.6\textwidth}
        \centering
        \includegraphics[width=\linewidth]{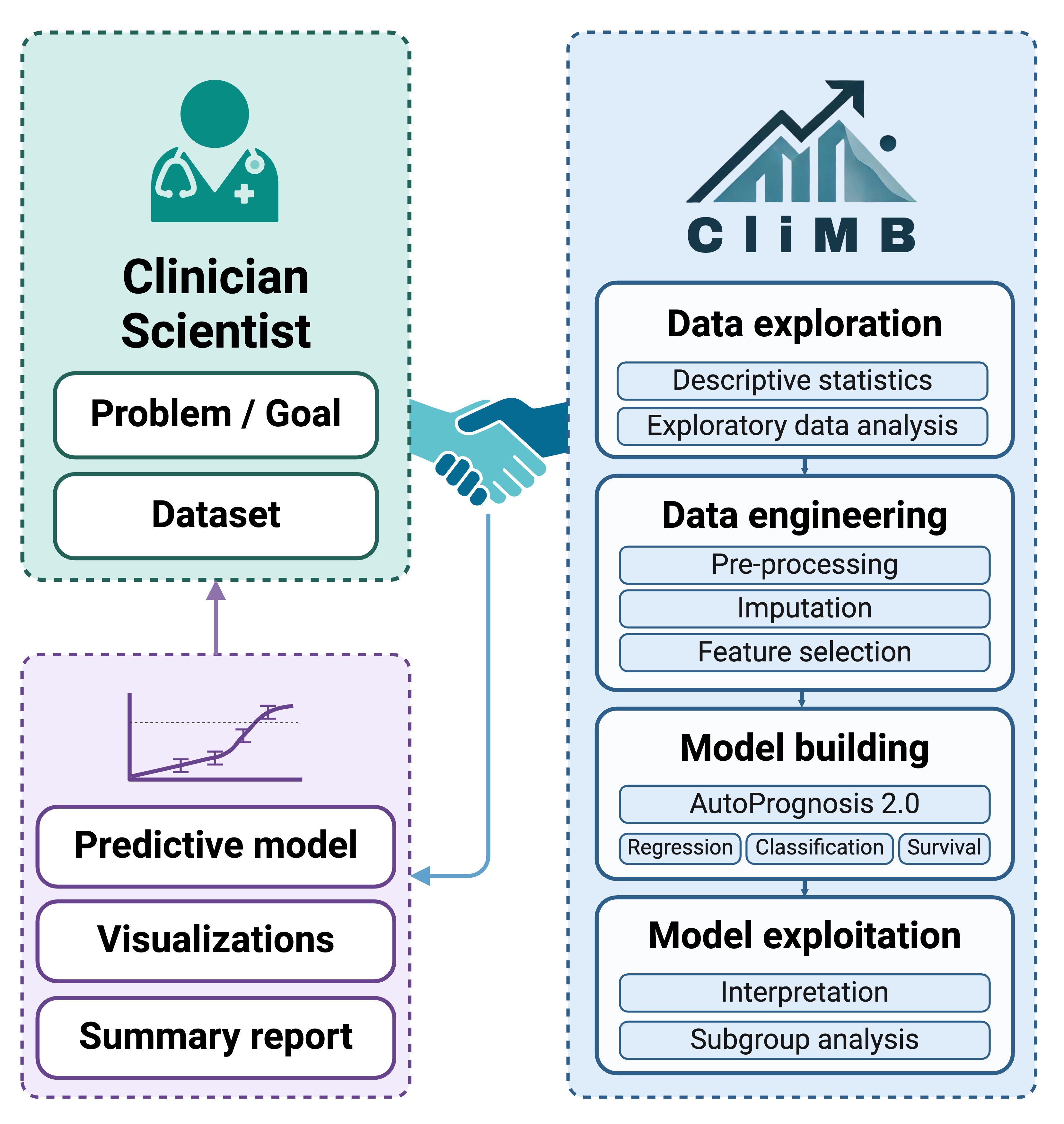}
    \end{minipage}%
    \hspace{0.02\textwidth} 
    \begin{minipage}{0.35\textwidth}
        \caption{\textbf{An AI-enabled partner for clinical predictive model building: \textit{CliMB}.} Clinician scientists turn to \textit{CliMB} with a predictive problem and real-world data. \textit{CliMB} guides the clinician scientist through all phases of the data science pipeline with robust planning and SOTA tools, including the AutoML pipeline AutoPrognosis 2.0 \cite{Imrie2023-wp} alongside data-centric and interpretability tools (see Table~\ref{table:available_tools}). The clinician scientist and \textit{CliMB} partner to generate a predictive model, visualizations, and a summary report of the methodology. \vspace{0.5em} \\ \scriptsize{Figure created with Biorender.com}}
        \label{General Idea Figure}
    \end{minipage}
\end{figure}

In this article, we introduce the \textit{\textbf{Cli}nical predictive \textbf{M}odel \textbf{B}uilding partner (CliMB)}, a no-code chatbot that empowers clinician scientists to build predictive models using only natural (specialist) language. \textit{CliMB} supports the entire medical data science pipeline, adhering to the requirements outlined in section \ref{requirements}. Using \textit{CliMB}, clinician scientists can build a predictive model from real-world clinical data in one conversation. We demonstrate the advanced capabilities of \textit{CliMB} through (1) sessions in which clinician scientists use \textit{CliMB} to build predictive models with their own medical datasets, (2) simulated modeling sessions to systematically compare \textit{CliMB} against a \baseline, largely similar to ChatGPT \cite{OpenAI2023-gd}, and (3) a survey among a diverse group of clinicians.

The proof-of-concept implementation of \textit{CliMB} is available as open-source software on GitHub (\href{https://github.com/vanderschaarlab/climb}{https://github.com/vanderschaarlab/climb}). Comprehensive project documentation is provided at \href{https://climb-ai.readthedocs.io/en/latest/}{https://climb-ai.readthedocs.io/en/latest/}. Detailed installation instructions can be accessed at \href{https://climb-ai.readthedocs.io/en/latest/installation.html}{https://climb-ai.readthedocs.io/en/latest/installation.html}, and a quickstart guide is available here: \href{https://climb-ai.readthedocs.io/en/latest/quickstart.html}{https://climb-ai.readthedocs.io/en/latest/quickstart.html}.

\section{Methods}

\subsection{Meeting the human-centric AI/ML partner requirements}\label{section:methods:requirement_fulfillment}

\begin{table}[h!]
    \centering
    \begin{tabularx}{\textwidth}{p{4cm}|X}
        \toprule
        \textbf{Requirement} & \textbf{Solution in \textit{CliMB}} \\
        \midrule
        Holistic support & End-to-end support for medical data science pipeline with access to SOTA tools in every phase. \\
        \midrule
        Usability & Intuitive user interface with interactive chat and dashboard. \vspace{0.5em} \\
        & Automation of many basic tasks, with human feedback requested when necessary.\\
        \midrule
        Best practices & Adherence to a structured plan for creating predictive models aligned with data science best practices. \\
        \midrule
        Robustness & Advanced reasoning unit formalizes the AI/ML partner problem as a transparent episodic multi-armed bandit to ensure the above plan is followed (see section \ref{section:methods:reasoning_unit}). \vspace{0.5em}\\
        & Memory unit containing a working directory, structured plan, files, and logs. \vspace{0.5em}\\
        & Use of external SOTA data science tools (see Table \ref{table:available_tools}).\\
        \midrule
        Interpretability & Intuitive user interface with natural language and visual explanations. \vspace{0.5em}\\
        & Integrated data-centric tools for understanding the clinical dataset and post-hoc interpretability methods for model assessment (see Table \ref{table:available_tools}).\\
        \midrule
        Modularity & Easy-to-extend tool library and modular LLM component. \\
        \midrule
        Versatility & Access to diverse tools and classes of predictive models. \vspace{0.5em}\\
        & Flexible handling of user requests.\\
        \midrule
        Privacy & Projects are performed locally and on secure and encrypted Azure servers. \\
        \bottomrule
    \end{tabularx}
    \caption{\textbf{Overview of requirements and solutions in \textit{CliMB}.}}
    \label{overview of requirements and solutions}
\end{table}

We built \textit{CliMB} to fulfill the universal requirements for a clinician scientist-centric AI partner. \textit{CliMB} offers holistic support for end-to-end predictive modeling. Its user-friendly natural language interface is complemented by an interactive dashboard, which allows the user to easily view data and data transformations, figures, and progress on the structured plan. This structured plan for predictive modeling is aligned with data science best practices and includes all relevant subtasks (for details see Table \ref{tabplan}). For instance, \textit{CliMB} checks for data leakage - an issue that was regularly overlooked by clinician scientists before we added this subtask. \textit{CliMB}'s reasoning unit, in combination with the memory unit, and the use of external tools ensure robustness. The reasoning unit is largely responsible for adherence to the structured plan. The memory unit stores logs and files but also all generated and executed code. This information can be retrieved by the reasoning unit to improve robustness, but it can also be viewed by the user to enable transparency. External tools are leveraged to process and interpret data, automate model building, and provide post-hoc interpretability (see Table \ref{table:available_tools}). These tools, combined with \textit{CliMB}'s user interface ensure the interpretability and transparency of the pipeline. While the reasoning unit ensures that a structured plan is being followed, \textit{CliMB} can flexibly adapt to user requests and still ensure that all necessary steps of the plan will be followed. This, alongside access to diverse tools and classes of predictive models, underpins the versatility of \textit{CliMB}.

Finally, data privacy is a critical concern in healthcare that we aim to address. In this context, clinician scientists, as data holders, must act as data stewards to ensure that data usage aligns with legal and ethical standards.\cite{Price2019-zc} To decrease the risk of privacy violations, one approach is the \textit{ex ante} limitation of data access \cite{Terry2012-ui}. Beyond the performance limitations of \baseline~demonstrated in section \ref{exp_comparison_against_gpt4V}, using commercial chatbots like ChatGPT with health data poses data security risks \cite{Li2023-pf}. Effectively using ChatGPT and similar tools for medical data science requires disclosing patient data. However, OpenAI uses past user interactions to retrain their models, and these logs cannot be deleted \cite{Tayebi_Arasteh2024-ov}. Data access could be restricted in several ways of which we will discuss three: prompt-based guardrails, explicit user confirmation, and adjustments to the computing infrastructure. 

First, when relying on a third-party LLM provider like OpenAI, users may have some control over which information the LLM accesses. For example, the LLM can be instructed to avoid accessing individual data records and instead focus on summary statistics. However, this ``safeguard'' is weak and can be easily bypassed by LLM errors or by the user inadvertently instructing it to access the data.

Second, a mechanism can be implemented to require explicit user confirmation for any content sent to the LLM. While this approach could prevent the unintentional release of private data, it would demand significant user effort. Moreover, both of the above options restrict data access, and thus limit the LLM's ability to perform complex reasoning based on the clinical data.

Therefore, we decided to focus on the computing infrastructure. OpenAI's GPT-4 can be deployed on trusted servers like Azure, which provide private instances with multiple layers of data security and encryption, fully compliant with GDPR \cite{UnknownUnknown-as}. This approach ensures that confidential information is neither shared with OpenAI nor used to retrain GPT-4. We chose this solution for the initial version of \textit{CliMB}. In the future, GPT-4 could also be substituted by other LLMs (e.g., the open source LLM LLaMA \cite{UnknownUnknown-og}) that can be run on private servers (e.g., hospital-owned) or even local machines, depending on the specific requirements of the clinician scientist. In addition to private deployment of the LLM, the dataset file is stored locally, and all code execution performed by \textit{CliMB} occurs on the user's machine. More information is provided in section \ref{section:appendix:data_privacy}.

Table \ref{overview of requirements and solutions} summarizes how \textit{CliMB} fulfills the universal requirements for a human-centric AI/ML partner.

\subsection{Design of CliMB}
\label{design-climb}

\begin{figure}[h!]
    \centering
    \includegraphics[width=1\linewidth]{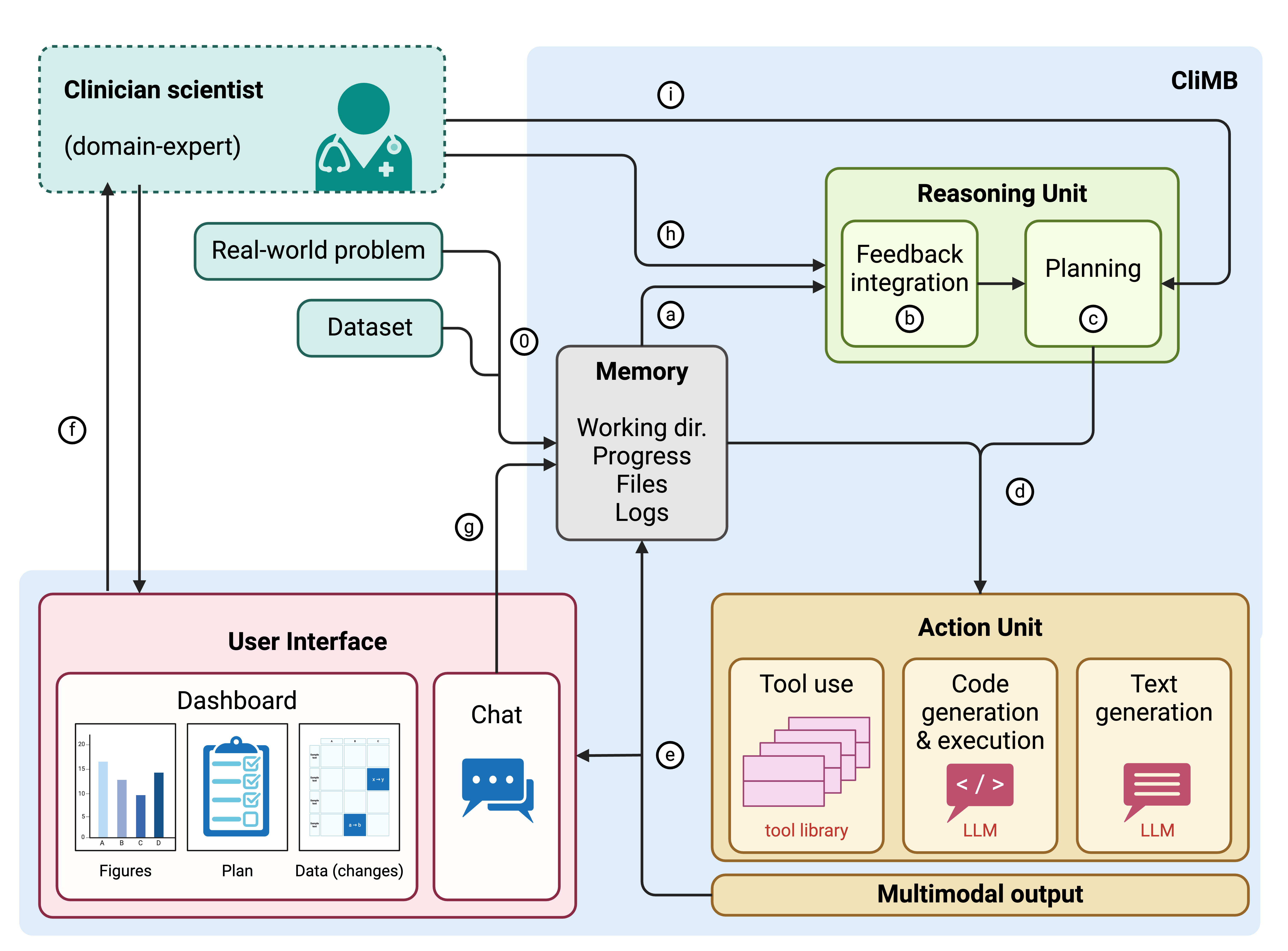}
    \caption{\textbf{Design of CliMB.} The information graph highlights the flow of information within \textit{CliMB}. The process begins with the user describing their clinical problem and uploading a corresponding medical dataset (0). The memory unit stores a working directory for tools, continuously updated plans tracking overall progress, files, and logs of the entire user interaction (g). The reasoning unit receives information from the memory (a), user (h), and through self reflection, and integrates this feedback (b) to facilitate subsequent planning (c). The action unit executes the plan (d) and generates multimodal outputs, which are stored in the memory and displayed via the user interface (e). The clinician scientist interacts with the user interface (f) and, at the end of an episode (see section \ref{section:methods:reasoning_unit}), validates and concludes the current phase (i). \vspace{0.5em} \\ \scriptsize{Figure created with Biorender.com}}
    \label{CliMB Design Figure}
\end{figure}

In this section, we describe the information flow through \textit{CliMB}.
\textit{CliMB} adheres to data science best practices, guiding users through their projects. These projects can be divided into the following phases: (1) data exploration, (2) data engineering, (3) model building, and (4) model exploitation \cite{De_Bie2022-eu}. 

The objective is to build predictive models for the user’s problem using real-world medical datasets. Therefore, the clinician scientist plays a central role within \textit{CliMB}’s human-in-the-loop framework. Other key components include the reasoning unit, the action unit, the memory, and the user interface (see Figure \ref{CliMB Design Figure}).

The reasoning unit receives information from the user, external tools invoked by the action unit, and through self-reflection. The reasoning unit (detailed in section \ref{section:methods:reasoning_unit}) integrates this input, updates the project plan, and assigns tasks to the action unit. It ensures robustness in following the planned stages necessary for predictive modeling. Planning assistance is necessary to make ML accessible but has been underexplored compared to execution assistance \cite{Gu2024-th}. This likely stems from the fact that many AutoML tools seem to cater mostly to data scientists wanting to enhance their workflow, who are already familiar with standard data science project plans. In contrast, our focus is on clinician scientists who require more support in structuring their predictive modeling efforts.

The action unit is modular by design, currently capable of performing three classes of actions: (1) tool use, (2) code generation/execution, and (3) text generation. These actions are invoked based on tasks assigned by the reasoning unit. For code and text generation, a SOTA LLM is integrated, which can be modularly replaced with more advanced models or those offering enhanced privacy features. Additionally, the action unit has access to SOTA data science tools in three categories: (a) data-centric tools (for phases 1, 2, and 4 of the data science pipeline), (b) AutoML model-building tools (for phase 3), and (c) post-hoc interpretability tools (for phase 4). These tools are summarized in Table \ref{table:available_tools}. Notably, the action unit’s modularity allows for seamless integration of a wide range of existing and emerging tools to expand the capabilities of \textit{CliMB}.

\begin{table}[h!]
    \centering
    \begin{tabularx}{\textwidth}{p{4cm}|X|p{2cm}}
        \toprule
        \textbf{Tool class} & \textbf{Available tools} & \textbf{Data science phases} \\
        \midrule
        Data-centric & Descriptive statistics, Exploratory data analysis (EDA), HyperImpute \cite{Jarrett2022-qt}, Feature selection \cite{Remeseiro2019-zx}, DataIQ \cite{Seedat2022-al} & (1), (2), (4) \\
        \midrule
        Model building & AutoPrognosis 2.0 \cite{Imrie2023-wp} (Regression, Classification, Survival analysis) & (3)\\
        \midrule
        Post-hoc interpretability &  SHAP explainer \cite{Lundberg2017-gq}, Permutation explainer \cite{Breiman2001-bf}, AutoPrognosis subgroup analysis \cite{Imrie2023-wp} & (4)\\
        \bottomrule
    \end{tabularx}
    \caption{\textbf{Overview of tools available to the action unit in CliMB.}}
    \label{table:available_tools}
\end{table}

The multimodal output produced by the action unit is routed through the memory unit to the reasoning unit for reflection. Following an iterative process between the reasoning and action units, the reasoning unit directs the action unit to deliver the multimodal output to the user interface and store it in the memory.

The user interface features a dynamic chat-like component alongside a stable dashboard that provides real-time updates on the current plan, highlights any dataset changes, and displays figures and other files.

Above, we have outlined the general flows of information, which over the course of the data science project, are followed to build a predictive model. In section \ref{exp_illustrative_sessions}, we present illustrative real-world clinical predictive modeling projects completed by clinician scientists using \textit{CliMB}. The design of \textit{CliMB} is summarized in Figure \ref{CliMB Design Figure}.

\subsection{Reasoning unit}\label{section:methods:reasoning_unit}

\textit{CliMB}’s reasoning unit formalizes the clinician scientist–AI partner collaboration problem as a transparent episodic multi-armed bandit \cite{Tekin2015-sm}. This formalism enables the description of both the sequence of decisions made by \textit{CliMB} to build a predictive model with its accompanying visualizations and report, and the feedback mechanism informing the next decision. The division of the end-to-end application into planned subtasks is made possible by the episodic nature of the model.

\begin{figure}[h!]
    \centering
    \includegraphics[width=\linewidth]{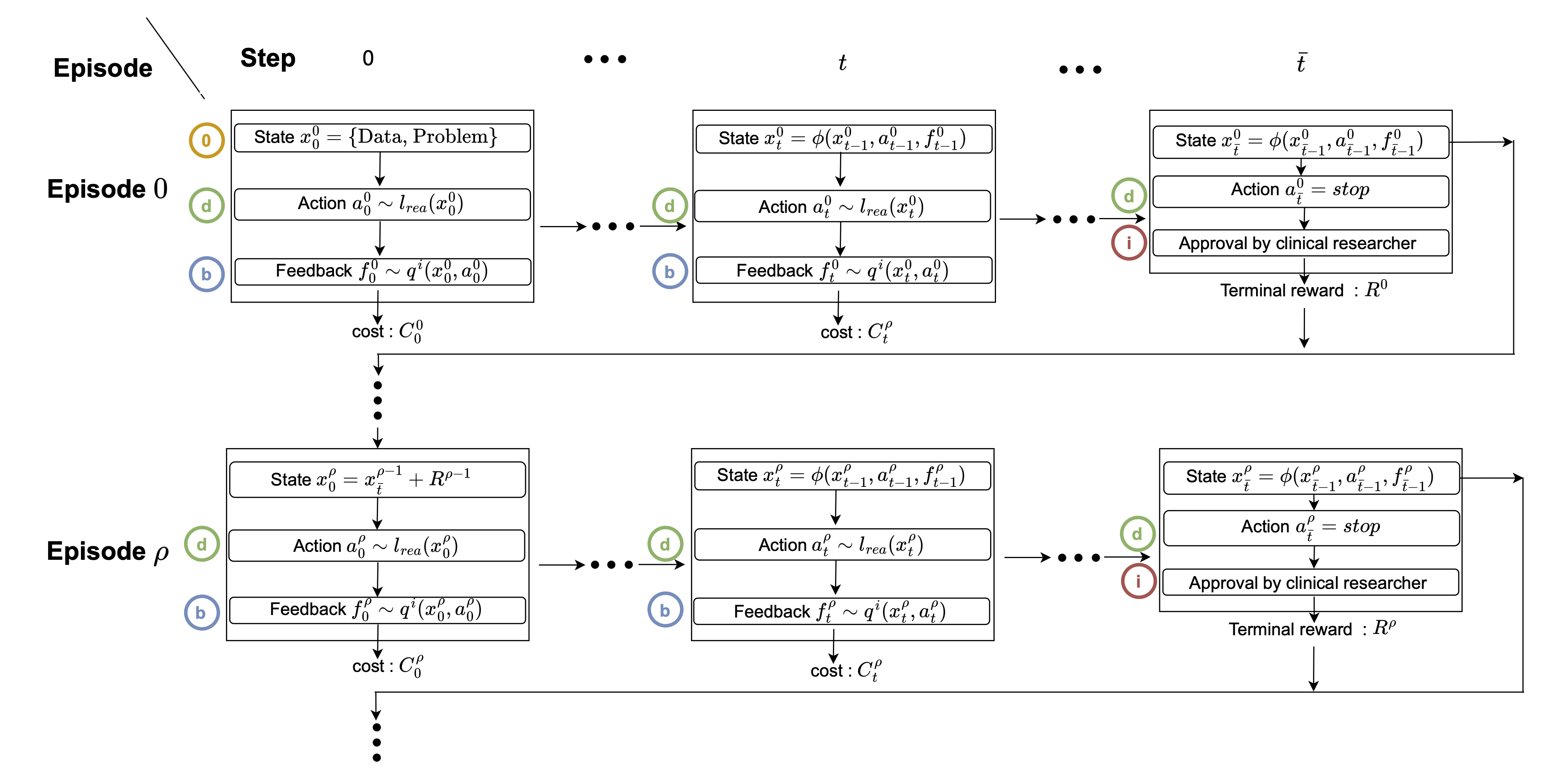}
    \caption{\textbf{The reasoning unit modeled as an episodic multi-armed bandit.} The integration of this unit in the information graph detailed in Figure \ref{CliMB Design Figure} is as follows: \textcolor{orange_bis}{\circled{0}} Initial objective given by the user, \textcolor{blue_bis}{\circled{b}} Feedback mechanism, \textcolor{green_bis}{\circled{d}} Calling of actions, \textcolor{red_bis}{\circled{i}} Validation by the user.}
    \label{fig:method}
\end{figure}

This formalism is used to structure, through in-context learning, a locally deployed pre-trained language model $l_{rea}$ (see Figure \ref{fig:method}).  Let $\mathcal{T}$ be the vocabulary of tokens used by $l_{rea}$. Therefore, texts generated by  $l_{rea}$ belong to $\mathcal{T}^*$.

We define a fixed set $\mathcal{E}$  of project stages, or episode types, composing the ``medical data science pipeline.'' These episode types or subtasks can be categorized into the following categories: ``alignment check'', ``data exploration'', ``data engineering'', ``model building'', and ``model exploitation''. All subtasks are provided in appendix \ref{appendix:structured_plan}. During an episode $\rho$, \textit{CliMB} takes a sequence of actions $(a^{\rho}_0,a^{\rho}_1,...,a^{\rho}_t)$ in $\bar{\mathcal{A}}$. These actions can be: (i) continuation actions $\mathcal{A}$, which progress toward the final result (model, visuals, report), or (ii) a terminal action $stop$, ending the current episode and requesting human verification. To prevent infinite iterations, we set a maximum number of actions $l_{max}$. If the terminal action is not selected before this limit, then $a^{\rho}_{l_{max}} = stop$.

The action $a^{\rho}_t \in \bar{A}$ is selected based on the state $x^{\rho}_t$, which includes the last feedback $f^{\rho}_{t-1}$ : $a^{\rho}_t = l_{rea}(x^{\rho}_t)$. States and feedback are defined as follows:

\interfootnotelinepenalty=10000
\textbf{States.} The states $ \mathcal{X} $ describe the current progress of the task. As the task may change depending on the episode, a state $ x $ can be a textual representation ($x \in \mathcal{T}^*$) of various objects such as a dataset in the ``Data engineering'' phase or a model in the ``Model building'' phase. The dynamics of the evolution of $ x $ caused by the action selected by \textit{CliMB} is given by a state transition function $ \phi \footnote{Unlike in the original episodic multi-armed bandits formalism, we constrain the state transition function to be time-homogenous.} : \mathcal{X} \times \mathcal{A} \times \mathcal{F} \mapsto \mathcal{X} $. The last feedback $ f^{\rho}_t \in \mathcal{F} $ is concatenated to the new state $ x^{\rho}_{t+1} $.
For each episode, the initial state $ x^{\rho}_0 $ corresponds to the last state of the previous episode $ x^{\rho-1}_t $ along with its associated terminal reward $ R^{\rho} $ (see paragraph below), if $ \rho > 0 $. If $ \rho = 0 $, the initial state corresponds to the data and problem provided at the beginning by the clinician scientist.

\textbf{Feedback.} The feedback provides interpretations of the current state of the task with additional information, guiding \textit{CliMB} in selecting the next action. The feedback is expressed in natural language: $ \mathcal{F} \subset \mathcal{T}^* $. The feedback is produced after executing an action by querying a feedback source $q^i$ : $f^{\rho}_t \sim q^i(x^\rho_t,a^\rho_t)$. \newline There are three sources for feedback ($i \in \{1,2,3\}$): 
\begin{itemize}
     \item[\ding{172}] External tools
    \item[\ding{173}] Self-reflection of the LLM
    \item[\ding{174}] Clinician scientist
\end{itemize}
We prioritize feedback from sources (i) or (ii) to maximize \textit{CliMB}'s autonomous behavior, resorting to human interaction only when necessary. Thus, the cost of an action $ c_{x,a} \in \{0,1\} $ for $ a \in \mathcal{A} $ and $ x \in \mathcal{X}$ is defined as $ c_{x,a} = 1 $ if it queries the clinician scientist and $ c_{x,a} = 0 $ if it queries another source. The cost of the stop action is always 0.

\textbf{Terminal reward and AI/ML partner objective.} In \textit{CliMB}, the terminal action $ stop $ is tied to an assessment by the clinical researcher of the episode process and resulting state. The episodic terminal reward, $ R^{\rho} $ for the episode $ \rho $, is binary ($ R^{\rho} \in \{0,1\} $), provided by the clinician scientist, corresponding to a quality and trustworthiness check. Thus, the episodic reward is 1 if the clinician scientist gives their approval and 0 otherwise. \textit{CliMB}'s objective is to maximize terminal rewards, leading to the completion of the subtasks required to obtain the final output (model, visuals, report), while minimizing action costs, i.e., human interactions.

\textbf{Robust Planning.}  
A plan is defined as a set of subtasks $\mathcal{E}$. The subtask attributed to episode $\rho$ is denoted $e_\rho$. The plan is completed when all subtasks $\hat{e} \in \mathcal{E}$ are achieved, meaning there exists an episode $\rho \in \mathbb{N}$ where $e_\rho = \hat{e}$ and $R_\rho = 1$. This guarantees robustness as it ensures that all steps of the data science pipeline are followed according to best practices. Although $\mathcal{E}$ is fixed, the order of completing subtasks is adjusted during the process, based on the problem, current progress, and feedback: $e_{\rho} = l_{CliMB}(x^{\rho-1}_{\hat{t}},R^{\rho-1})$. The plan $\mathcal{E}$ is described in a dashboard section which is regularly updated to track the progress of the different subtasks.

\textbf{Tools available.} Let $\mathcal{A}^- $ be the set of actions available to a general-purpose LLM, including planning, text generation, ideation, etc. In \textit{CliMB}, we provide the AI/ML partner with a code interpreter and multiple machine learning tools necessary for acting on the state. Without these coding tools, $\mathcal{A}^- $ alone would not suffice for performing the required actions and iterating autonomously based on the tool's results. Depending on the episode type, the set of tools available $\mathcal{G} $ may vary but can be categorized as shown in Table \ref{table:available_tools}.

Thus, the set of continuation actions $\mathcal{A} $ is enriched by the available tools $\mathcal{G} $, increasing the range of potential actions for the AI partner: $\mathcal{A} = \mathcal{A}^- \cup \mathcal{G} $.

\textbf{Scalability.}  We propose \textit{CliMB} as a modular framework, adaptable to various tasks depending on the diversity of episode types defined and the quality of available tools $ \mathcal{G} $. Since this paper focuses on generating risk prediction models for clinician scientists, the episode types correspond to the subtasks required to achieve this goal, and the tools provided are data science tools. However, the framework could be applied to other tasks, and the quality of results could be enhanced by improving the set of tools.

\textbf{Summary of the clinician scientist interactions.} \textit{CliMB}'s reasoning unit has two types of interactions with the clinician scientist. First, the edge {\circled{h}} of the information flow graph presented in Figure \ref{CliMB Design Figure}: the feedback mechanism allows querying the researcher for intermediary approval of decisions or for additional expert knowledge, helping \textit{CliMB} progress on the task. This type of interaction is minimized when possible by choosing other feedback sources, as it is associated with a non-zero cost. Second, the edge {\circled{i}} of the information flow graph in Figure \ref{CliMB Design Figure}: the validation step at the end of each episode provides the clinician scientist with the opportunity to assess the quality of the results and the trustworthiness of the process. For their decisions, the clinician scientist can rely on the information provided by the dashboard and through the chat.

\subsection{End-to-end Pipeline}
\textit{CliMB}’s ability for robust planning, as described in section \ref{section:methods:reasoning_unit}, is realized by its reasoning unit. \textit{CliMB} guides the user through a medical data science project by following an explicit (yet adaptable) plan aligned with best practices. Each necessary stage \cite{De_Bie2022-eu} is ensured to be completed unless the user explicitly decides to skip it. In this section, we outline the standard pipeline, which can be adapted to meet specific user needs. A detailed overview of the plan is available in Table \ref{tabplan}.

\begin{itemize}
    \item[\faCheckCircleO] \textbf{Alignment check.} \textit{CliMB} verifies whether the hardware is suitable and whether the dataset can be loaded and used. Background information on the dataset and the clinician scientist’s research question(s) is confirmed to ensure alignment between \textit{CliMB}’s capabilities and the user’s requirements.
    \item[\ding{172}] \textbf{Data exploration.} \textit{CliMB} performs detailed exploratory data analysis, generates descriptive statistics, and creates informative figures for each variable. The meaning of each column and the expected data type is confirmed with the user.
    \item[\ding{173}] \textbf{Data engineering.} This phase includes data type conversion and the explicit discussion and handling of missing values. Additionally, any changes requested by the user are performed. For full transparency, data changes are highlighted within the dashboard. Feature selection is conducted using random-forest-based algorithms.
    \item[\ding{174}] \textbf{Model building.} Model building is optimized using SOTA AutoML risk prediction software (AutoPrognosis 2.0 \cite{Imrie2023-wp}). Currently, classification, regression, and time-to-event (survival) analysis can be performed.
    \item[\ding{175}] \textbf{Model exploitation.} In this phase, model performance is evaluated, and depending on the project type, additional insights are provided to the user. Interpretability methods are employed to assess feature importance \cite{Lundberg2017-gq,Breiman2001-bf} and, for classification tasks, stratify samples (easy, ambiguous, hard) \cite{Seedat2022-al}. Once all steps are completed and the clinician scientist is satisfied with the results, a summary report of the study is generated.
\end{itemize}

\noindent We show the breakdown of \textit{CliMB} project stages' tasks and subtasks in appendix \ref{appendix:structured_plan}, and an illustration of the advantages of \textit{CliMB}'s approach in robustly planning an AI-enabled data science project in appendix \ref{appendix:planning_illustration}.

\subsection{Hardware and Software}

\textit{CliMB} is implemented in Python, with backward compatibility up to Python 3.8. The software is compatible with all major operating systems: Windows, Linux, and macOS. The \texttt{miniconda} \cite{Anaconda-Inc2016-qi} package and environment manager is used to facilitate code and tool execution. Automatic installation of PyPI \cite{UnknownUnknown-ry} (\texttt{pip}-installable) packages in a dedicated code execution environment is implemented.

The experiments were carried out using the \texttt{gpt-4-0125-preview} model with a 128,000-token context window via an Azure OpenAI Service \cite{UnknownUnknown-ox} deployment. However, all OpenAI and Azure OpenAI Service provisions of the \texttt{gpt3.5-turbo}, \texttt{gpt4(-turbo)}, and \texttt{gpt4o} class models are compatible through the Python \texttt{openai} library \cite{UnknownUnknown-qu}. Compatibility with other commercial and open-source LLMs is available through simple extension, hence \textit{CliMB} is not limited to the models used here. The packages required for \textit{CliMB} tools, such as AutoPrognosis 2.0 \cite{Imrie2023-wp} or HyperImpute \cite{Jarrett2022-qt}, are installed as part of the \textit{CliMB} setup.

The hardware used for developing and running \textit{CliMB} was a workstation running Ubuntu 24.04, powered by an Intel Core i9-10900K 10-core, 20-thread CPU, 64 GB of system memory, and an NVIDIA GeForce RTX 3090 GPU (24 GB VRAM). We estimate the minimum system specifications for running \textit{CliMB} to be: a CPU with at least 4 cores, 16 GB of system RAM; optionally, a GPU with at least 4 GB of memory is advantageous for running some of the deep learning models. However, it should be noted that the appropriate computing requirements depend significantly on the size of the user's dataset.

\section{Results}

\subsection{Illustrative end-to-end sessions with clinician scientists}\label{exp_illustrative_sessions}
\begin{figure}[htbp]
    \centering
    \vspace{-0.7cm}
    \includegraphics[width=1\textwidth]{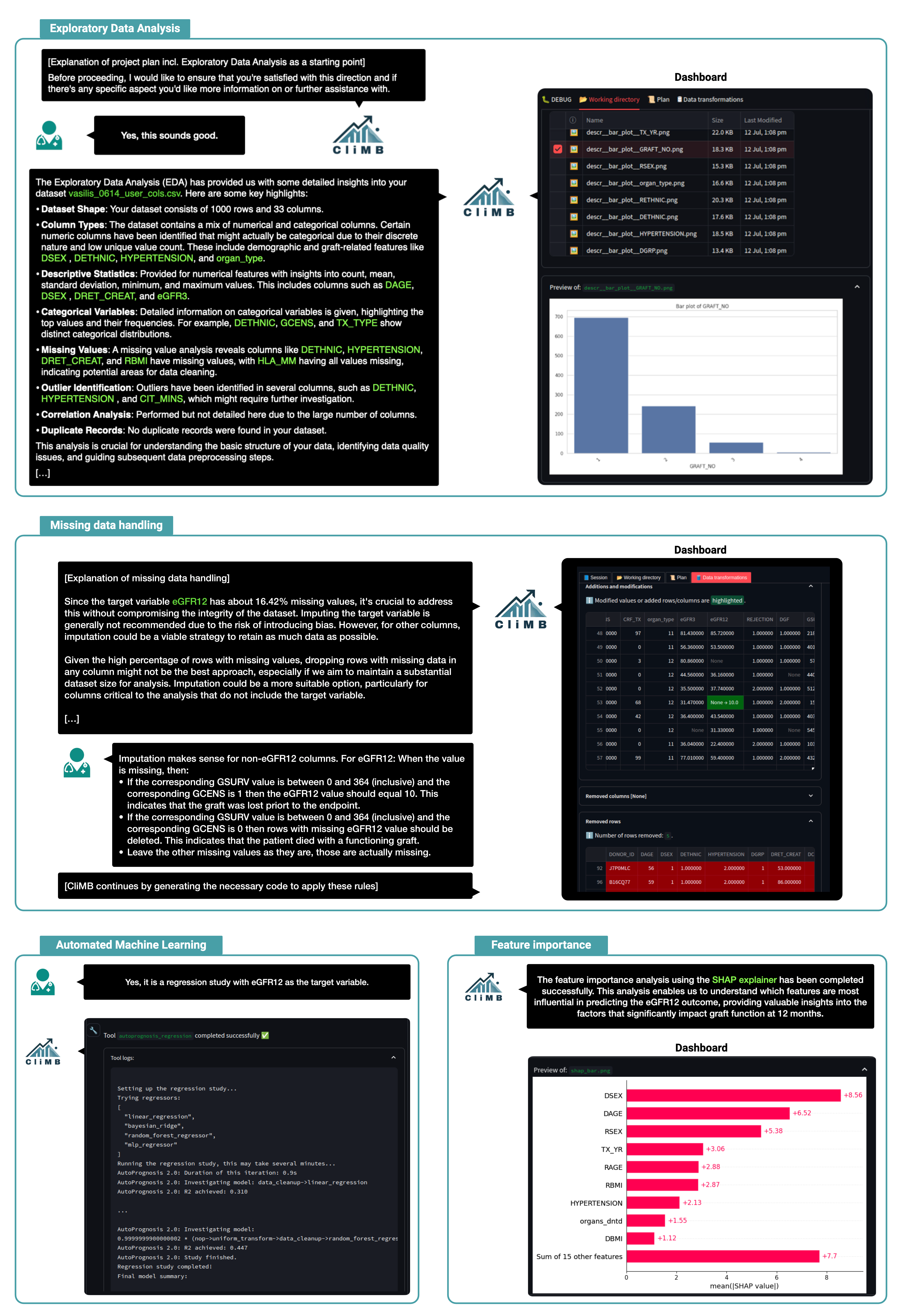}
    \vspace{-0.7cm}
    \caption{\textbf{Illustrative usage of CliMB.} Snippets of interactions during a full end-to-end model-building session are provided from all four phases of the medical data science project: (1) data exploration (shown: Exploratory Data Analysis), (2) data engineering (shown: data transformation), (3) model building (shown: AutoPrognosis 2.0 Survival Analysis \cite{Imrie2023-wp}), and (4) model exploitation (shown: SHAP explainer \cite{Lundberg2017-gq} for interpreting feature importance). \textit{Note:} illustrative examples shown do not necessarily correspond to the sessions used in experiments. \vspace{0.5em} \\ \scriptsize{Figure created with Biorender.com}}
    \label{Fig: CliMB Illustrative Usage}
\end{figure}

We tested \textit{CliMB}’s capabilities to assist real-world clinician scientists in creating predictive models from their medical datasets. Predictive models for various use cases and diseases were created during these sessions. Illustrative logs are included in the appendix. We further include a video of a full \textit{CliMB} session here \url{https://youtu.be/76XuR0K3F5Y}. Figure \ref{Fig: CliMB Illustrative Usage} provides snippets from this session in which a transplant surgeon built a predictive model for graft function at 12 months. These snippets highlight just a few of the versatile capabilities of \textit{CliMB}.

\subsection{Systematic comparison against \baseline} \label{exp_comparison_against_gpt4V}

\begin{table}[h!]
    \centering
    \begin{tabularx}{\textwidth}{p{5cm}|X|X}
        \toprule
        \textbf{Feature} & \textbf{CliMB} & \textbf{\Baseline} \\
        \midrule        
        Flexible adaptation to user prompts & \checkmark & \checkmark \\
        \midrule
        Robust planning & \checkmark & \texttimes \\
        \midrule
        Advanced memory & \checkmark & \texttimes \\
        \midrule
        Access to SOTA external tools & \checkmark & \texttimes \\
        \midrule
        Informative dashboard & \checkmark & \texttimes \\
        \midrule
        Privacy features & \checkmark & \texttimes \\
        \bottomrule
    \end{tabularx}
    \caption{\textbf{Comparison of features between CliMB and \baseline.}}
    \label{comparison of features}
\end{table}

We systematically compare \textit{CliMB} with a \baseline~software system. The latter is implemented as a \texttt{gpt4-turbo} chat interface with code execution capabilities with self-reflection. The LLM, its parameters, and the code execution environment implementation are equivalent in both the baseline and \textit{CliMB} to ensure a fair comparison. Python code execution with self-reflection, access to (at least some) \texttt{pip}-installable libraries and a working directory, as implemented in the baseline, appears to be similar to ChatGPT's \cite{OpenAIUnknown-un} approach to code execution and data analysis as of August 2024, based on empirical observation of the ChatGPT online portal's code generation behavior (for instance, generated code and feedback from exceptions can be confirmed by assessing the content of the ``Analyzing...'' snippet). We do, however, note that the exact implementation details of ChatGPT are not disclosed by OpenAI. We opt for implementing a baseline system rather than using ChatGPT online as (i) this allows us to make the baseline and \textit{CliMB} implementations directly comparable, and (ii) this allows the use of the Azure OpenAI Service, which was necessary for data privacy when working with real-world clinical data.

From the outset, \baseline~lacks several crucial capabilities that \textit{CliMB} integrates. These are summarized in Table \ref{comparison of features}.
To systematically compare \textit{CliMB}’s capabilities with those of \baseline, standardized sessions were conducted in which a simulated clinician scientist built predictive models from the same clinical dataset using both \textit{CliMB} and \baseline. 

\begin{samepage}
The following assumptions were made for the simulated clinician scientist persona:

\begin{itemize}
     \item[\ding{172}] No deeper knowledge of data science, machine learning or coding
    \item[\ding{173}] Knowledge of their clinical field
    \item[\ding{174}] Knowledge of the variables in their dataset
    \item[\ding{175}] A clear idea of the predictions they want to make
\end{itemize}
\end{samepage}

\textbf{Experiment setup.} To create a realistic scenario, we used a real-world clinical dataset that has not been published before and is therefore not part of the corpus GPT-4 was trained on. This dataset contains information about kidney transplants performed in UK hospitals. Variables in the dataset are categorized into donor characteristics, recipient characteristics, and graft characteristics (including survival). During the interaction, the user requested two dataset-specific steps: (1) removal of specific columns and (2) subgroup analysis by donor sex.

The objective during the end-to-end runs with both chatbots was to create a predictive model for graft function at 12 months (eGFR12). This task was performed with \textit{CliMB} and \baseline~(5 replicates each). Full logs of these interactions were then analyzed.

The evaluation metrics were computed on a held-out test set that was not used during model development. After completing the training session on the initial dataset using either the baseline approach or the \textit{CliMB} system, the same data preprocessing steps were applied to the separate test dataset. The final model generated by either method was then used to make predictions on the preprocessed test set, which were subsequently evaluated.

\textbf{Failures in various categories.} Failures were categorized and are presented in Table \ref{table: Failure category comparison}. While both \textit{CliMB} and \baseline~completed all end-to-end runs, \baseline~exhibited more frequent failures across diverse categories and stages over the five runs.

\begin{table}[h!]
\centering
\caption{\textbf{Failures of CliMB and \baseline~in various categories.} The table lists the proportion of runs in which a failure type occurred.}
\label{table: Failure category comparison}
\begin{tabularx}{\textwidth}{p{4cm}|X|X|X}
\toprule
\textbf{Failure category} & \textbf{Stage} & \textbf{\Baseline} & \textbf{CliMB} \\
\midrule
Did not finish & N/A & 0/5 & 0/5 \\
\midrule
EDA partially failed & Data exploration & 1/5 & 0/5 \\
\midrule
Did not save the models (unless prompted) & Model building & 3/5 & 0/5 \\
\midrule
Did not provide an opportunity for user to review or transform data features for their needs & Data engineering & 2/5 & 0/5 \\
\midrule
Imputed the target variable without checking with the user & Data engineering (missing values) & 4/5 & 0/5 \\
\midrule
Dropped rows with any missing values, resulting in dropping most samples in the dataset & Data engineering (missing values) & 1/5 & 0/5 \\
\midrule
Did not use cross validation & Model building & 4/5 & 0/5 \\
\midrule
Subgroup analysis done by retraining instead of using existing model & Model exploitation & 2/5 & 0/5 \\
\bottomrule
\end{tabularx}
\end{table}

\textbf{Identification of dataset specific issues.} In addition, we quantified whether each chatbot would identify two issues specific to the real-world dataset used: (1) the inclusion of ID columns, which could lead to model overfitting and hamper interpretability; and (2) the presence of certain columns which, if included in the model, would result in data leakage because these values are not available at the study baseline time (these fields include a second measure of graft function at 3 months, graft survival time and censoring variables, rejection, delayed graft function, and cause of graft failure). In all instances (5/5), \baseline~missed both issues, while \textit{CliMB} missed issue 1 in 0/5 runs and issue 2 in only 1/5 runs.

\textbf{Planning failures.} To investigate the planning capabilities of \textit{CliMB} and \baseline, we assessed whether each chatbot completed all tasks it had planned in each phase. This evaluation included only the tasks planned by the chatbot itself during the session. \textit{CliMB} demonstrated robust planning, consistently completing all planned tasks in each stage. In contrast, \baseline~frequently failed to complete tasks it had planned (see Table \ref{table: planning failure comparison}).

\begin{table}[h!]
\centering
\caption{\textbf{Planning failures.} The table lists the proportion of runs in which the respective chatbot failed to complete all its planned tasks for the specific stage.}
\label{table: planning failure comparison}
\begin{tabularx}{\textwidth}{p{5.5cm}|X|X}
\toprule
\textbf{Stage} & \textbf{\Baseline} & \textbf{CliMB} \\
\midrule
Data exploration and engineering & 3/5 & 0/5 \\
\midrule
Model building & 0/5 & 0/5 \\
\midrule
Model exploitation & 1/5 & 0/5 \\
\bottomrule
\end{tabularx}
\end{table}

\textbf{Code generation and execution.} The success of code generation and execution was measured by the average number of exceptions encountered per run. \textit{CliMB} encountered significantly fewer exceptions (0.4 ± 0.4) compared to \baseline~(4.6 ± 2.3) (unpaired t-test, t(8) = 3.202, p = 0.013).

\textbf{Model performance.} Finally, we measured the performance of the predictive model from each session using four metrics: mean squared error (MSE), root mean squared error (RMSE), mean absolute error (MAE), and $R^2$. When a data leak was not removed during the data phases, the metrics were computed after manually excluding the leak feature. \textit{CliMB} performed significantly better than \baseline~across all metrics, as summarized in Table \ref{table: performance metrics}.

\begin{table}[h!]
\centering
\caption{\textbf{Model performance.} Comparison between \textit{CliMB} and \baseline~across key metrics (MSE, RMSE, MAE, R²). The mean and standard deviation for each metric are reported. \textit{CliMB} significantly outperforms \baseline~across all metrics. Statistical significance is indicated by asterisks based on unpaired t-tests (following outlier removal where applicable): \(^{**}\) p \textless 0.01, \(^{***}\) p \textless 0.001, \(^{****}\) p \textless 0.0001. †: 1 outlier identified and removed using the ROUT method (Q = 1\%) before performing unpaired t-tests.}
\label{table: performance metrics}
\begin{tabularx}{\textwidth}{p{3.2cm}|X|X|X|X}
\toprule
 & \textbf{MSE} & \textbf{RMSE} & \textbf{MAE} & \textbf{R²} \\
\midrule
\textbf{\baseline} & 433.7 (173.6)\textsuperscript{†} & 20.55 (3.78)\textsuperscript{†} & 15.50 (3.69)\textsuperscript{†} & 0.148 (0.208) \\
\midrule
\textbf{CliMB} & 286.3 (5.7)\textsuperscript{***} & 16.81 (0.09)\textsuperscript{****} & 12.03 (0.46)\textsuperscript{***} & 0.485 (0.014)\textsuperscript{**} \\
\bottomrule
\end{tabularx}
\end{table}

\subsection{Clinician feedback}\label{subsection:Clinician_feedback}
To get a first impression of \textit{CliMB}'s perceived usefulness in real-world application, we gathered feedback from clinicians at the \textit{Cambridge AI in Medicine Summer School} \cite{noauthor_undated-ip}. As part of the summer school curriculum, video demonstrations (see appendix \ref{subsection:school_vids}) were shown to participants, initially without any additional commentary, allowing them to focus fully on the videos. In these videos, the same type of predictive model was built from the same dataset twice: once with \baseline, once with \textit{CliMB}. While the goal of the summer school session was to demonstrate, how \textit{CliMB} truly enables the attendees to build predictive models, there was no indication as to which tool was being used in either demonstration, effectively resulting in a blinded evaluation. Following the demonstrations, participants voted in a survey to compare \textit{CliMB} against \baseline. Consent was obtained (see appendix section \ref{subsection:consent_form} for the full consent form and information provided to participants). 

\begin{figure}[h!]
    \centering
    \includegraphics[width=1\linewidth]{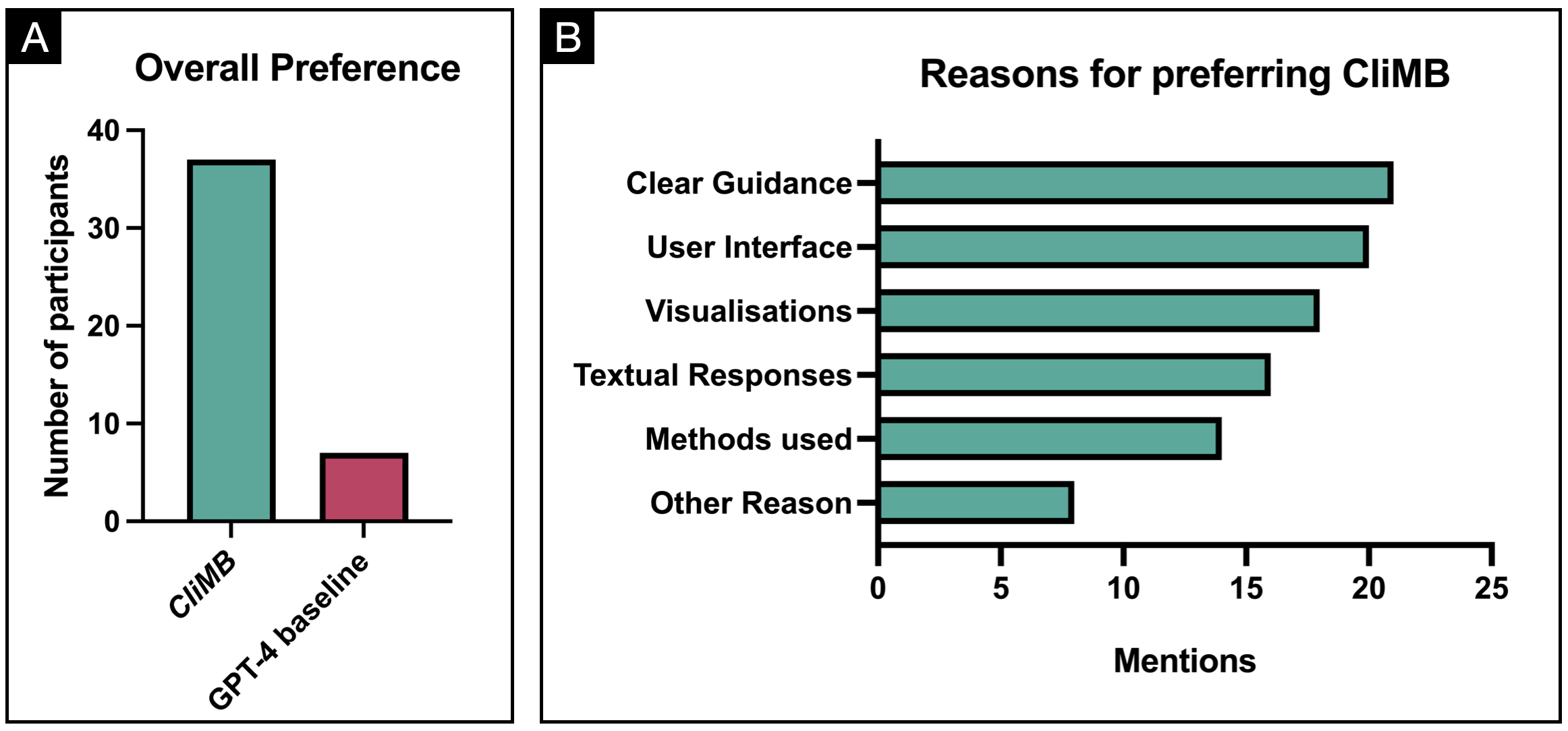}
    \caption{\textbf{Clinician preferences in a direct comparison of \textit{CliMB} and \baseline.} \textbf{A)} Overall, 37/45 participants preferred \textit{CliMB} to \baseline. \textbf{B)} The bar plot on the right summarizes the reasons mentioned by participants who preferred \textit{CliMB}.}
    \label{Figure:combined_clinician_feedback_plots}
\end{figure}

Clinicians from diverse career stages participated - from medical students to consultants/attending physicians. Overall, about 82\% of these clinicians preferred \textit{CliMB} over \baseline. The most frequent reasons for choosing \textit{CliMB} over \baseline~included its clear guidance and user interface (see Figure \ref{Figure:combined_clinician_feedback_plots}).

\section{Discussion}

\textit{CliMB} is a first-of-its-kind solution to the domain expert-AI-conundrum for clinical research. It meets the universal requirements outlined in section \ref{requirements}, as explained in section \ref{section:methods:requirement_fulfillment}. 

\subsection{CliMB outperforms \baseline}

In section \ref{exp_comparison_against_gpt4V}, we demonstrated that \textit{CliMB} consistently outperforms \baseline~across multiple dimensions. First, \textit{CliMB} typically avoids a variety of potential failures during different stages of the data science project. It also exhibits a strong capability to identify dataset-specific issues, such as potential data leaks or risks of overfitting. By successfully flagging these issues in nearly every instance, \textit{CliMB} minimizes risks that could compromise the validity of the models. In contrast, \baseline~consistently overlooked these issues, which raises concerns about its reliability in real-world applications.

These differences reflect the advantages of the advanced reasoning unit of \textit{CliMB}. \textit{CliMB} demonstrated robust planning by ensuring that all planned steps of the workflow were completed. The planned steps themselves follow data science best practices. In contrast, \baseline~frequently failed to complete even the tasks it had planned itself, which could lead to invalid analyses, downstream failures, and missed quality checks.

Moreover, \textit{CliMB}’s code generation and execution capabilities were markedly superior, with significantly fewer exceptions encountered during runs.

Finally, the models generated by \textit{CliMB} consistently outperformed those created by \baseline~across key performance metrics that reflect higher accuracy and predictive power.

Most importantly, \textit{CliMB} is preferred by clinicians as demonstrated in section \ref{subsection:Clinician_feedback}. Interestingly, the most common reason for this preference was the clear guidance provided by \textit{CliMB}, which it can provide thanks to its structured plan. These statements also highlight the demand for planning assistance.

To sum up, \textit{CliMB}’s infrastructure, with its robust planning and SOTA tools, delivers a more dependable and effective solution for medical data science. \textit{CliMB} represents a significant advancement over \baseline, all while incorporating data privacy.

\subsection{Limitations}
Our work has several limitations that are outlined below.
\begin{itemize}
    \item[\ding{172}] While the need for technical skills is mostly eliminated, clinician scientists who want to build and use predictive models still require other skills and knowledge. Specifically, they need a basic understanding of ML and a thorough understanding of their dataset to avoid overreliance on the system.
    \item[\ding{173}] The predictive models created with \textit{CliMB} require further validation and must be tested for clinical effectiveness and safety before deployment \cite{Longhurst2024-cn}. Therefore, clinicians who wish to deploy predictive models in practice need a proficient level of AI expertise. By that, we mean an understanding of the ethical implications of integrating AI in medicine and the ability to critically evaluate the predictive models created with \textit{CliMB}.
    \item[\ding{174}] Finally, \textit{CliMB} should undergo more systematic evaluation in collaboration with clinician scientists from various disciplines. This process should encompass all types of predictive problems and utilize datasets of varying sizes with different dataset-specific challenges.
\end{itemize}

\subsection{Future vision}
In this work, we present a core technology that is both versatile, making it applicable across various specialties, and modular, allowing it to scale with ease. \textit{CliMB} can be augmented with additional tools to handle multi-modal data. Moreover, future versions can expand to include machine learning methodology beyond risk prediction (e.g., causal machine learning for CATE estimation \cite{Feuerriegel2024-ua}) or handle competing risks during survival analysis \cite{Alaa2017-el}. In addition to expanding the core version, we envision more specialized versions of \textit{CliMB}-like tools. To provide an example, these could include multi-modal chatbots offered by registries to interact with the registry data.

\textit{CliMB} was built with the vision of an ideal AI/ML partner for clinical risk prediction in mind (see section \ref{section:methods:requirement_fulfillment}). As a result, \textit{CliMB} democratizes predictive modeling in medicine. It is one of the first realizations of the vision of a collaborative design that integrates the complementary strengths of human expertise and AutoML methodologies, as articulated by Lindauer and colleagues \cite{Lindauer2024-th}. \textit{CliMB} extends this vision by also incorporating data-centric and post-hoc interpretability methods (see Table \ref{table:available_tools}) and by offering robust planning assistance. This development enables a larger and more diverse group of clinician scientists to build predictive models for the real-world problems they face in patient care. On one hand, it bridges the gap between powerful existing solutions and their application in medicine. On the other hand, by scaling the use of ML in real-world settings, new problems will be identified by domain experts and relayed to the ML community, which can focus on finding technical solutions to unsolved problems. \textit{CliMB} facilitates a dialogue that will accelerate progress toward reality-centric AI.

Ultimately, \textit{CliMB} brings the community closer to realizing AI’s full potential in healthcare. Our goal is to empower clinician scientists through \textit{CliMB} to build accurate, trustworthy, and interpretable predictive models that have a tangible positive impact on patient care. We want to empower clinician scientists to build machine learning that matters \cite{Wagstaff2012-qx}.

\section{Acknowledgments}
We thank Andres Floto, Tim Oosterlinck, Andriy Melnik, Tom Callender, Janet Allen, Rachel Smith, Will Brown, and Courtney Kremler for their invaluable clinical knowledge and feedback on \textit{CliMB} and Andrew Rashbass, Nabeel Seedat, Jiashuo Liu, Nicolás Astorga, Alicia Curth, Paulius Rauba, Robert Davis, Claudio Fanconi, Paul Schubert, and Favian Bauer for helpful discussions. This work was supported by Microsoft's Accelerate Foundation Models Academic Research initiative. TS acknowledges scholarships from the Medical Faculty at Heidelberg University and the German Academic Scholarship Foundation. VK acknowledges funding as a Paul I. Terasaki Scholar (G106170). 

\newpage

\bibliography{main.bib}

\newpage

\input{appendix}

\end{document}

%% file: appendix.tex
\section{Appendix}

\subsection{Software architecture}
The software architecture of \textit{CliMB} is founded on the following design principles:
\begin{enumerate}
    \item \textbf{LLM, tool, and environment-agnostic.} \textit{CliMB} is not tied to a particular LLM in its design, set of tools (locally executable, predefined code snippets with a human readable documentation), or code execution environment. The LLM used can be changed via simple extension of the \texttt{Engine} class, new tools can be defined as needed, and the code execution environment is compartmentalized, flexible and managed via \texttt{conda}. A notable limitation currently is that only Python code execution environments are supported, but this choice is based on the observation that the most extensive set of ML and data science libraries are available primarily in Python.
    \item \textbf{Modular.} The \textit{CliMB} software design is modular, broadly following the SOLID \cite{Martin2000-bv} design principles. The main concepts in the architectures are: \texttt{Engine}, which defines how reasoning and action units work via the LLM; \texttt{Message}, which stores the details of each interaction step (between the LLM, user, tools, and code execution); \texttt{UI} which defines a user interface compatible with the rest of the architecture; and the \emph{code execution environment} logic.
    \item \textbf{Extensible.} The two preceding points make \textit{CliMB} extensible - the key components of the system can be built upon in a straight-forward fashion. In addition, a structured plan that defines the details of the project to be performed between the LLM and the user can be customized easily, via a JSON configuration file.
\end{enumerate}


\subsection{Data privacy considerations}\label{section:appendix:data_privacy}
The approach to data privacy considerations used in \textit{CliMB} is further elaborated in this section. This approach was chosen in order to ensure a level of data privacy that is compatible with the requirements of certain non-public datasets. The use of each dataset utilized in the experiments reported was approved with the clinician scientist acting as the data steward. In addition, no personally identifiable fields were present in any of the datasets, and all identification reference fields (e.g. database IDs that may correspond to data records of a specific patient, procedure, treatment etc.) were replaced with randomly generated values prior to use with \textit{CliMB}. The following key points characterize \textit{CliMB}'s current approach to data privacy:
\begin{enumerate}
    \item \textbf{Local storage of data.} All dataset files (original or modified in any way by \textit{CliMB}) are stored locally on the user's machine. Hence, the data files are never uploaded to any third-party (LLM provider's, or other) servers.
    \item \textbf{Local code execution.} All code execution performed by \textit{CliMB}, either through code generation or predefined tool invocation occurs locally on the user's machine. Hence, no working directory files of any kind (including saved predictive models, image files, tool output artifacts etc.) leave the user's machine.
    \item \textbf{Use of Azure OpenAI Service as LLM provider.} Azure OpenAI Service, deployed within the researchers' Azure tenancy, was used to access LLM text generation. In addition to customer data ownership control in Azure \cite{UnknownUnknown-as}, Azure OpenAI Service has additional data privacy considerations \cite{ChrisHMSFTUnknown-dq}. Namely, the user-LLM text interactions (composed of input prompts and output completions):
    \begin{itemize}
        \item are not available to other Azure customers,
        \item are not available to OpenAI,
        \item are not used to improve OpenAI models,
        \item are not used to train, retrain, or improve Azure OpenAI Service foundation models,
        \item are not used to improve any Microsoft or third party products or services without the customer's permission or instruction.
    \end{itemize}
    While Azure OpenAI may store some data that is necessary for stateful features to function \cite{ChrisHMSFTUnknown-dq}, in the case of \textit{CliMB} this is minimal; while it may apply to the functionality of tool calling, in the form of the temporary tool call IDs, which we use, we do not utilize the major stateful features such as the Assistant API, Files API, or vector store. Furthermore, any data that is stored:
    \begin{itemize}
        \item is stored at rest in the Azure OpenAI resource in the customer’s Azure tenant, within the same geography as the Azure OpenAI resource (for all datasets used in the experiments, the geography was set to correspond to that of the dataset and the clinician scientist);
        \item was double encrypted at rest, by default with Microsoft's AES-256 encryption;
        \item can be deleted by the customer at any time.
    \end{itemize}
    The Azure subscriptions hosting the OpenAI resources used for the experiments have also been set up under ``modified abuse monitoring'' exception \cite{ChrisHMSFTUnknown-dq}, which additionally entirely prevents temporary storage of the prompt history and ensures no human review of the data.
\end{enumerate}

\newpage
\subsection{Structured plan}
\label{appendix:structured_plan}
This section provides additional details of the structured plan used in \textit{CliMB} in order to improve project planning robustness. The plan is displayed in Table \ref{tabplan}. In addition to the task and subtask names, the plan contains a more detailed description for each. The ``subtask selection'' field indicates if the subtask is to be issued always (``mandatory''), or depending on certain conditions checked against the project memory (``conditional'') -- conditional subtasks have a corresponding condition specification, e.g. ``applicable to survival analysis''.

\begin{table}[h!]
\centering
\small

\renewcommand{\arraystretch}{1.1}

\begin{tabular}{p{2.3cm}|>{\RaggedRight\arraybackslash}p{2.5cm}|p{7.5cm}|p{1.5cm}}

\hline
\textbf{Project stage} & \textbf{Task name} & \textbf{Subtask name} & \textbf{Subtask selection} \\
\hline

\multirow{6}{=}{Alignment check} 
& \multirow{3}{=}{Getting the environment ready} 
& Upload data file & mandatory \\
& & Check hardware & mandatory \\
& & Check data file can be loaded & mandatory \\
\cline{2-4}
& \multirow{3}{=}{Getting information from the user} 
& High-level information from the user & mandatory \\
& & Experiment setup and research question from the user & mandatory \\
& & Assess data suitability and tool support & mandatory \\
\hline

\multirow{5}{=}{Data exploration} 
& \multirow{5}{=}{Exploratory Data Analysis} 
& Exclude/keep columns & mandatory \\
& & Perform EDA & mandatory \\
& & Generate descriptive statistics & mandatory \\
& & Warn about small sample size if necessary & conditional \\
& & Show Kaplan-Meier plot & mandatory \\
\hline

\multirow{11}{=}{Data engineering} 
& Data processing - before missing 
& Column background information & mandatory \\
\cline{2-4}
& \multirow{4}{=}{Data processing - missing data handling} 
& Represent missing data as NaN & mandatory \\
& & Consider dropping columns with high missing values & mandatory \\
& & Consider dropping rows with missing values & mandatory \\
& & Impute missing values & conditional \\
\cline{2-4}
& \multirow{2}{=}{Data processing - after missing} 
& Discuss data preprocessing with the user & mandatory \\
& & Feature selection & mandatory \\
\cline{2-4}
& \multirow{4}{=}{Machine learning study pre-checks} 
& Confirm ML problem type & mandatory \\
& & Check time and event columns & conditional \\
& & Check for data leakage & mandatory \\
& & Check for irrelevant columns & mandatory \\
\cline{2-4}
\hline

\multirow{4}{=}{Model building}
& \multirow{3}{=}{Machine learning study} 
& Machine learning study - classification & conditional \\
& & Machine learning study - regression & conditional \\
& & Machine learning study - survival analysis & conditional \\
\cline{2-4}
& ML iteration 
& Iterate the machine learning study with the user & mandatory \\
\hline

\multirow{3}{=}{Model exploitation} 
& \multirow{3}{=}{Machine learning study exploitation} 
& Feature importance plots & mandatory \\
& & Insights on classification & conditional \\
& & Subgroup analysis & mandatory \\
\hline

End of Study 
& Finish up the project 
& Discuss the project and finish up & mandatory \\
\hline

\end{tabular}

\caption{\textbf{Structured project plan}. This table summarizes all stages, tasks, and subtasks.}
\label{tabplan}
\end{table}

\newpage
\subsection{Illustration of the planning process}
\label{appendix:planning_illustration}
\textit{CliMB}'s reasoning unit coordinates the project flow through iterative execution of the underlying subtasks. The structured plan defined for the project provides ``checkpoints'' that represent \emph{guardrails} (consider, ``can we proceed to the next stage of the workflow given the current state of the data?'') and \emph{best practices} (``the recommended approach in situation $X$ is $Y$'').

\begin{figure}[htbp]
    \centering
    \includegraphics[width=1\textwidth]{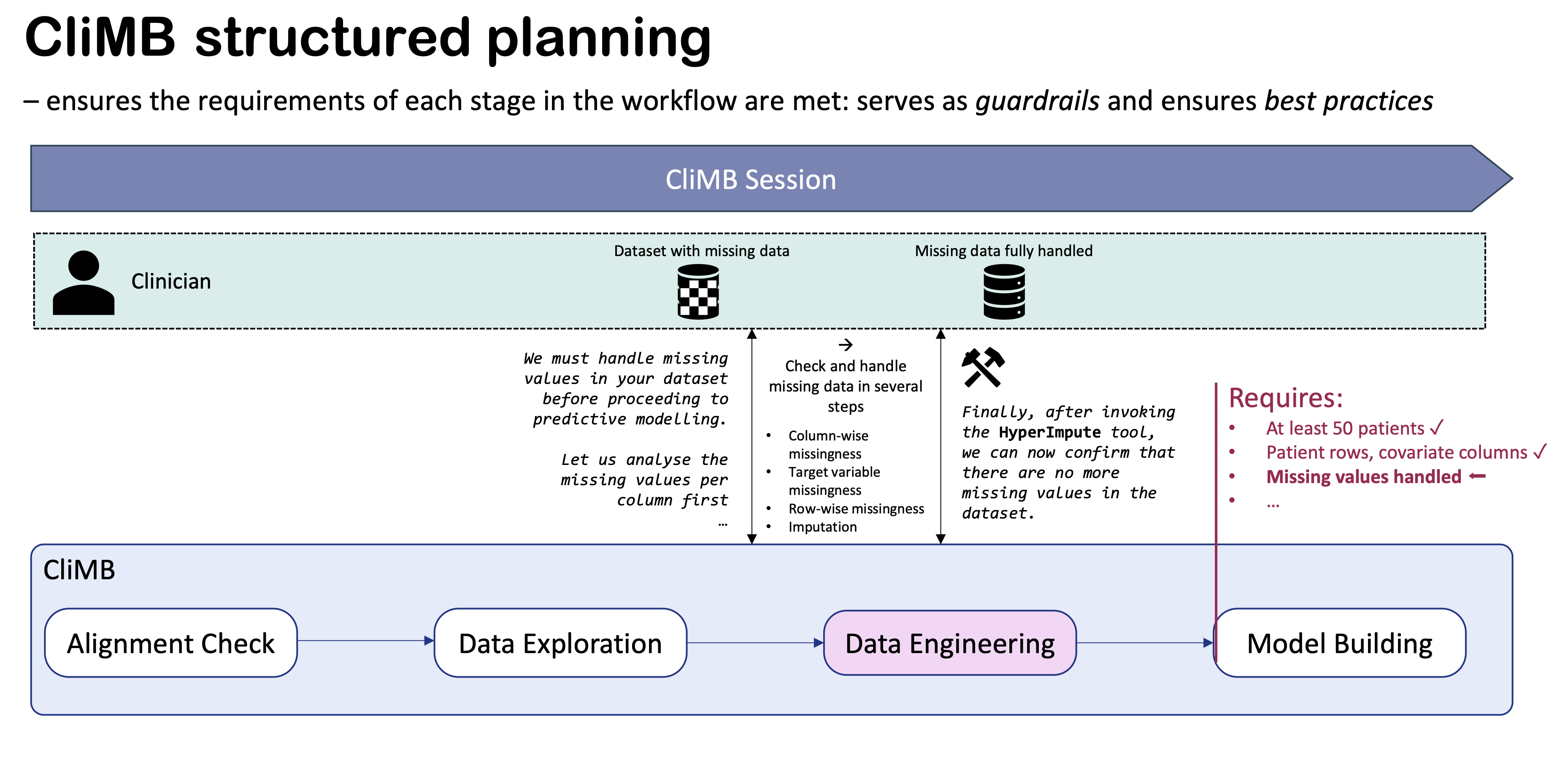}
    \caption{\textbf{\textit{CliMB} planning checkpoint illustration.} Illustration of the ``data engineering'' stage's checkpoint within the project flow. The reasoning unit ensures the dataset is evaluated for missing values before transitioning to the ``model building'' stage. The checkpoint involves several subtasks that ensure the dataset is prepared and complete for subsequent steps in the workflow.}
    \label{fig:plan_image}
\end{figure}

This is well-illustrated through the scenario shown in Fig \ref{fig:plan_image}. Given that the ``model building'' project stage requires, among other points, that there are no missing values present, the plan for the ``data engineering'' stage contains the appropriate missing data handling checkpoint. This manifests itself as an evaluation of the dataset for missing values and execution of several subtask actions (all informed by user-guided decisions through feedback): column-wise missingness evaluation, row-wise missingness evaluation, investigation of the target variable's missing values, and imputation. On the output of this process, the dataset is devoid of missing values, and it is safe to proceed to the subsequent ``model building'' work.

In the absence of this approach, we find that the stochastic nature of LLMs leads to frequent inconsistencies in the state of the project as it progresses (including the dataset, model, and other files in the working directory, as well as the actual tasks undertaken), and consequently machine learning  project best practices are sometimes not followed.

\newpage
\subsection{Interaction log examples}

\newcommand{\mydots}{\noindent \texttt{...}}
\newcommand{\user}{\textbf{User:}~}
\newcommand{\assistant}{\textbf{Assistant:}~}
\newcommand{\generatedCode}{\textbf{Generated code:}~}
\newcommand{\codeExecOut}{\textbf{Code execution output:}~}
\newcommand{\codeExecSuccess}{\newline \small{\texttt{
Code execution finished successfully \cmark~}}}
\newcommand{\codeExecFail}{\newline \small{\texttt{
Code execution failed \xmark~}}}

\newcommand{\toolExecOut}[1]{\textbf{Tool \texttt{{#1}} completed successfully \cmark}~}
\newcommand{\toolLogs}{\textbf{Tool logs:}~}
\newcommand{\toolOut}{\textbf{Tool output:}~}
\newcommand{\toolReport}{\textbf{The tool produced the following report:}~}

\newcommand{\mygap}{\noindent\rule{\textwidth}{1pt}}

\subsubsection{\Baseline~failure cases}

\subsubsection*{Example 1: incorrect missing value handling}

\noindent In the following example we see the failure case of \baseline~deciding to drop all missing values in order to avoid a code execution error (ultimately caused by missing values not being supported by the model to be fitted). This resulted in dropping the vast majority of the rows, leaving only 14 samples. The failure was not noticed by the system.

\mygap

\noindent \textbf{Summary of work up to this point:} An initial EDA was carried out, including some descriptive statistics, a correlogram; imputation of columns more than 80\% missing data was carried out; a number of columns were removed by user request. The dataset at this point has 39 columns and 1200 rows.

\mydots

\begin{assistantbox}
\assistant Would you like to start with Linear Regression, or do you have a preference for another regression algorithm?
\end{assistantbox}

\begin{userbox}
\user Let's start with that. I do not have data science expertise so I will need your guidance here.
\end{userbox}

\begin{assistantbox}
\assistant Absolutely, starting with Linear Regression is a sensible choice for its simplicity and interpretability. As we proceed, I'll guide you through each step, explaining the process and decisions along the way. \\

Here’s the plan for building and evaluating a Linear Regression model:

\begin{enumerate}
    \item Model Training: We'll use the training dataset (\texttt{X\_train.csv} and \texttt{y\_train.csv}) to train the Linear Regression model.
    \item Model Evaluation: After the model is trained, we'll use the testing dataset (\texttt{X\_test.csv} and \texttt{y\_test.csv}) to evaluate the model's performance. We'll look at metrics like RMSE (Root Mean Squared Error), which tells us how far off our predictions are from the actual values, on average; and (R\^{}2) (Coefficient of Determination), which gives us an idea of how well our model explains the variation in the target variable.
\end{enumerate}

Let's start by training the Linear Regression model with our training data. \\

\generatedCode
\begin{lstlisting}
import pandas as pd
from sklearn.linear_model import LinearRegression
from sklearn.metrics import mean_squared_error, r2_score
from math import sqrt

# Loading the training data
X_train = pd.read_csv('X_train.csv')
y_train = pd.read_csv('y_train.csv')

# Initializing the Linear Regression model
model = LinearRegression()

# Training the model with the training data
model.fit(X_train, y_train)

# Saving the trained model
import joblib
joblib.dump(model, 'linear_regression_model.pkl')

print("Linear Regression model trained and saved as 'linear_regression_model.pkl'.")
\end{lstlisting}

\end{assistantbox}

\begin{execbox}
\codeExecOut
\codeExecFail
\begin{lstlisting}[style=codeoutput]
Traceback (most recent call last):
  [...]
ValueError: could not convert string to float: 'Bw6'
\end{lstlisting}
\end{execbox}

\begin{assistantbox}
\assistant It appears we encountered an issue during the model training step because the dataset contains categorical variables, which Linear Regression cannot handle directly. Before training the model, it's essential to convert these categorical variables into a form that the model can process. One common approach is one-hot encoding, which converts categorical variables into a format of binary columns. \\

Here's how we can adjust our approach:

\begin{enumerate}
    \item Preprocess the Data: Perform one-hot encoding on categorical variables.
    \item Re-train the Model: With our data now fully numeric, we can train the Linear Regression model again.
\end{enumerate}

Let's start by preprocessing the training data with one-hot encoding for the categorical variables and then proceed to train the model. \\

\generatedCode
\begin{lstlisting}
import pandas as pd
from sklearn.compose import ColumnTransformer
from sklearn.linear_model import LinearRegression
from sklearn.metrics import mean_squared_error, r2_score
from sklearn.pipeline import Pipeline
from sklearn.preprocessing import OneHotEncoder
from sklearn.model_selection import train_test_split
from math import sqrt
import joblib

# Load the data
data = pd.read_csv('data_filtered.csv')

# Identify categorical columns
categorical_columns = data.select_dtypes(include=['object']).columns

# Preprocessing: One-hot encoding for categorical variables
preprocessor = ColumnTransformer(
    transformers=[
        ('cat', OneHotEncoder(handle_unknown='ignore'), categorical_columns)
    ],
    remainder='passthrough')

# Define the model
model = Pipeline(steps=[
    ('preprocessor', preprocessor),
    ('regressor', LinearRegression())
])

# Split the data into features (X) and target variable (y)
X = data.drop(columns=['eGFR12'])
y = data['eGFR12']

# Splitting the data back into training and testing sets
X_train, X_test, y_train, y_test = train_test_split(X, y, test_size=0.2, random_state=42)

# Fit the model
model.fit(X_train, y_train)

# Save the trained model
joblib.dump(model, 'linear_regression_model.pkl')

print("Linear Regression model trained with one-hot encoding and saved as 'linear_regression_model.pkl'.")
\end{lstlisting}

\end{assistantbox}

\begin{execbox}
\codeExecOut
\codeExecFail
\begin{lstlisting}[style=codeoutput]
Traceback (most recent call last):
  [...]
ValueError: Input X contains NaN.
LinearRegression does not accept missing values encoded as NaN natively. For supervised learning, you might want to consider sklearn.ensemble.HistGradientBoostingClassifier and Regressor which accept missing values encoded as NaNs natively. Alternatively, it is possible to preprocess the data, for instance by using an imputer transformer in a pipeline or drop samples with missing values. See https://scikit-learn.org/stable/modules/impute.html You can find a list of all estimators that handle NaN values at the following page: https://scikit-learn.org/stable/modules/impute.html#estimators-that-handle-nan-values
\end{lstlisting}
\end{execbox}

\begin{assistantbox}
\assistant The code execution encountered an issue due to the presence of NaN values in the data. This suggests that some steps in the data preprocessing phase were not fully addressed, particularly in dealing with all missing values before model training. \\

Given this situation, the most straightforward path forward involves ensuring that our dataset is free of any NaN values and that categorical variables are properly encoded. Since our initial approach to preprocessing within the model pipeline encountered challenges, I suggest we take a step back and manually preprocess the dataset to address NaN values and encode categorical variables before fitting the model again. \\

First, we'll handle NaN values across the entire dataset and encode categorical variables using one-hot encoding. After these preprocessing steps, we'll proceed with model training again. \\

Let's start by preprocessing the dataset. \\

\generatedCode
\begin{lstlisting}
import pandas as pd

# Load the filtered dataset
data = pd.read_csv("data_filtered.csv")

# Drop rows with any NaN values to ensure clean data for modeling
data_cleaned = data.dropna()

# Generating one-hot encodings for categorical variables
data_cleaned = pd.get_dummies(data_cleaned)

# Saving the preprocessed data
data_cleaned.to_csv("data_cleaned_preprocessed.csv", index=False)

print("Data cleaned and preprocessed. Saved as 'data_cleaned_preprocessed.csv'.")
\end{lstlisting}

\end{assistantbox}

\noindent Note the \texttt{df.dropna()} in the above generated code snippet causes this problem. The dataset at this point has 39 columns and only 14 rows.

\begin{execbox}
\codeExecOut
\codeExecSuccess
\begin{lstlisting}[style=codeoutput]
Data cleaned and preprocessed. Saved as 'data_cleaned_preprocessed.csv'.
\end{lstlisting}
\end{execbox}

\mydots

\noindent \textbf{Summary of work after this point:} The system and user moved on to feature importance analysis without realising the significant information loss.


\newpage
\subsubsection*{Example 2: failure to fully complete planned work}

\noindent In the following example, we observe the system failing to fully complete its intended work steps. In particular, during exploratory data analysis, an attempt to generate a correlation matrix is made, but abandoned due to an exception encountered. In addition, while the the work plan initially intends to perform feature selection, this ends up boiling down to simply listing the categorical and numeric variables and using them in the following predictive study stage, rather than using any feature selection method, or querying the user's domain knowledge.  

\mygap

\noindent \textbf{Summary of work up to this point:} The user's file was loaded, the columns were listed. The user then queried what the system suggests for next steps.

\mydots

\begin{userbox}
\user What do you suggest?
\end{userbox}

\begin{assistantbox}
\assistant Given the dataset and typical steps in a medical research project, I recommend proceeding with the following steps in order:

\begin{itemize}
    \item Data Cleaning: To handle missing values, which is crucial for ensuring the quality of the data for analysis. This includes deciding whether to impute missing values based on the distribution of the data or to drop columns/rows with a high percentage of missing values.
    
    \item Exploratory Data Analysis (EDA): To get a deeper understanding of the data, including the distribution of key variables, potential correlations between variables, and identifying any outliers. This step can help in formulating hypotheses for your study.

    \item Feature Selection: To identify the most relevant variables for the predictive modeling. This can be based on the results of the EDA, domain knowledge, or using statistical methods to find the most predictive features.

    \item Preparing for Machine Learning: This includes encoding categorical variables (if any), normalizing/scaling numeric variables to ensure that no variable dominates others due to its scale, and splitting the data into training and test sets.

    \item Machine Learning: Based on your research question, choosing a predictive task (classification, regression, survival analysis) and applying appropriate machine learning models. This also includes evaluating model performance and possibly iterating over different models or tuning parameters to improve results.
\end{itemize}

Would you like to start with Data Cleaning, move directly to Exploratory Data Analysis, or do you have another preference?
\end{assistantbox}

\begin{userbox}
\user Let's do as you suggested above.
\end{userbox}

\mydots

\noindent The system then proceeds with working with the user on removing highly missing columns user and imputing the missing target variable cases. 

\mydots

\begin{assistantbox}
\assistant The missing values in the \texttt{eGFR12} column have been imputed with the median, and columns with more than 50\% missing values have been removed. The dataset is now saved as \texttt{kidney\_prepared.csv}, ready for further analysis and modeling. \\

Next steps could include:

\begin{itemize}
    \item Exploratory Data Analysis (EDA): To further understand the distribution of the eGFR12 variable, relationships between features, and identify any potential outliers.
    \item Feature Selection: Identifying which variables are most relevant to predicting eGFR12.
    \item Preparing for Machine Learning: Encoding categorical variables, normalizing/scaling numeric variables, etc.
    \item Machine Learning: Building and evaluating a predictive model for eGFR12.
\end{itemize}

Would you like to proceed with Exploratory Data Analysis, or would you prefer to move directly into feature selection or model building?
\end{assistantbox}

\begin{userbox}
\user Yes, do that (EDA)
\end{userbox}

\begin{assistantbox}
\assistant Let's proceed with Exploratory Data Analysis (EDA) to better understand the dataset, especially focusing on the \texttt{eGFR12} variable, which is our target for prediction. EDA will help us to identify patterns, detect outliers, and understand the relationship between different variables. \\

For this EDA, we will:

\begin{enumerate}
    \item Describe the dataset statistically, including the central tendency, dispersion, and shape of the dataset's distribution.
    \item Visualize the distribution of \texttt{eGFR12} to understand its spread and identify any potential outliers.
    \item Visualize correlations between \texttt{eGFR12} and other variables to identify potential predictors.

We'll start with the statistical description and distribution visualization of \texttt{eGFR12}.
\end{enumerate}

\generatedCode
\begin{lstlisting}
import pandas as pd
import matplotlib.pyplot as plt
import seaborn as sns

# Load the prepared data
data = pd.read_csv("kidney_prepared.csv")

# Statistical description of the dataset
print("Statistical description of the dataset:")
print(data.describe())

# Distribution of eGFR12
plt.figure(figsize=(10, 6))
sns.histplot(data['eGFR12'], kde=True)
plt.title('Distribution of eGFR12')
plt.xlabel('eGFR12')
plt.ylabel('Frequency')
plt.savefig('eGFR12_distribution.png')
plt.close()

# Correlation matrix heatmap
plt.figure(figsize=(12, 10))
correlation_matrix = data.corr()
sns.heatmap(correlation_matrix, annot=False, cmap='coolwarm')
plt.title('Correlation Matrix Heatmap')
plt.savefig('correlation_matrix.png')
plt.close()

print("Distribution of eGFR12 visualized in 'eGFR12_distribution.png'.")
print("Correlation matrix heatmap saved as 'correlation_matrix.png'.")
\end{lstlisting}

\end{assistantbox}

\begin{execbox}
\codeExecOut
\codeExecFail
\begin{lstlisting}[style=codeoutput]
Statistical description of the dataset:
            DONOR_ID         DAGE  ...  PAST_DIABETES  PAST_HYPERTENSION
count    1200.000000  1200.000000  ...    1162.000000        1153.000000
mean    64055.581667    44.575833  ...       1.036145           1.193408
std     13818.714125    16.002531  ...       0.186730           0.395142
min     47359.000000     5.000000  ...       1.000000           1.000000
25%     54447.250000    34.000000  ...       1.000000           1.000000
50%     59765.500000    47.000000  ...       1.000000           1.000000
75%     69865.250000    57.000000  ...       1.000000           1.000000
max    111574.000000    77.000000  ...       2.000000           2.000000

[8 rows x 46 columns]

Traceback (most recent call last):
  [...]
ValueError: could not convert string to float: 'A2'
\end{lstlisting}
\end{execbox}

\begin{assistantbox}
\assistant The statistical description of the dataset was successful, providing insights into the central tendencies and dispersions of numerical variables. However, generating the correlation matrix heatmap failed due to the presence of non-numeric (string) data in the dataset, which cannot be directly correlated using the method attempted. \\

Given this, we successfully visualized the distribution of \texttt{eGFR12}, which should help in understanding its spread and identifying any potential outliers. The distribution visualization has been saved as \texttt{eGFR12\_distribution.png}. \\

Moving forward, we can:

\begin{itemize}
    \item Focus on numerical features for correlation analysis with \texttt{eGFR12}, excluding non-numeric columns to avoid errors.
    \item Proceed with feature selection and preparation for machine learning, considering only the variables relevant for predicting \texttt{eGFR12}.
\end{itemize}

Would you like to attempt a revised correlation analysis excluding non-numeric columns, or would you prefer to move on to the next steps in the analysis?
\end{assistantbox}

\begin{userbox}
\user Proceed to with the option you think makes more sense
\end{userbox}

\noindent Here we notice that the system moves on to the next steps without fixing the correlation matrix, citing the error encountered as the reason.

\begin{assistantbox}
\assistant Given the goal of predicting \texttt{eGFR12} and the error encountered with non-numeric data in the correlation analysis, it makes sense to proceed with feature selection and preparation for machine learning, focusing on variables relevant for predicting \texttt{eGFR12}. This approach will allow us to concentrate on preparing the data for modeling without further delay. \\

Steps for proceeding:

\begin{enumerate}
    \item Feature Selection: Identify numerical variables and potentially relevant categorical variables for predicting \texttt{eGFR12}. We'll consider variables that are likely to influence kidney function based on domain knowledge or previous studies.
    \item Data Preparation for Machine Learning:
    \begin{itemize}
        \item Encode Categorical Variables: Convert categorical variables into a format that can be provided to machine learning models, using techniques such as one-hot encoding.
        \item Normalize/Scale Numerical Variables: Ensure numerical variables are on a similar scale to improve model performance and convergence.
    \end{itemize}
\end{enumerate}

Let's start with feature selection, focusing on numerical variables and identifying key categorical variables for encoding. We'll prepare a list of features to be included in the model based on their potential relevance to \texttt{eGFR12}. \\

\generatedCode
\begin{lstlisting}
import pandas as pd

# Load the prepared data
data = pd.read_csv("kidney_prepared.csv")

# Identify numerical columns for potential inclusion
numerical_cols = data.select_dtypes(include=['int64', 'float64']).columns.tolist()

# Identify categorical columns for potential inclusion and encoding
categorical_cols = data.select_dtypes(include=['object']).columns.tolist()

# Print identified columns
print("Numerical Columns:")
print(numerical_cols)
print("\nCategorical Columns:")
print(categorical_cols)

# Note: This is an initial identification step. Further selection may be based on domain knowledge, correlation with 'eGFR12', or other statistical analyses.
\end{lstlisting}
\end{assistantbox}

\begin{execbox}
\codeExecOut
\codeExecFail
\begin{lstlisting}[style=codeoutput]
Numerical Columns:
['DONOR_ID', 'DAGE', 'DSEX', [...], 'ORGANS_TXD', 'PAST_DIABETES', 'PAST_HYPERTENSION']

Categorical Columns:
['Donor_First_A_Broad', 'Donor_Second_A_Broad', 'Donor_Second_A_Split', [...], 'Recip_First_DQ_Split', 'Recip_Second_DQ_Broad', 'recip_first_Bw']
\end{lstlisting}
\end{execbox}

\noindent We see that here, despite the system outlining that ``this is an initial identification step'' for feature selection, it proceeds directly to the dataset preparation without further action.

\begin{assistantbox}
\assistant The dataset contains a mix of numerical and categorical columns. Given our goal of predicting \texttt{eGFR12}, which is a numerical variable, we've identified the following columns for potential use in our predictive model:

\begin{itemize}
    \item Numerical Columns: These include donor and recipient demographic information, medical history, and pre- and post-transplant measurements. These variables are directly usable in regression models after scaling.
    \item Categorical Columns: These primarily consist of genetic markers and other categorical medical data. These will require encoding to be used in predictive modeling.
\end{itemize}

Next Steps:

\begin{enumerate}
    \item Encode Categorical Variables: We will convert categorical variables into a format suitable for machine learning models. Given the nature of the categorical data, mainly consisting of genetic markers, one-hot encoding is a suitable approach as it transforms categorical data into a format that can be provided to ML models without imposing ordinality.
    \item Normalize/Scale Numerical Variables: This step is crucial to ensure that all numerical variables contribute equally to the model's predictions. We'll scale the numerical variables to have a mean of 0 and a standard deviation of 1.
    \item Split the Data: Before training models, we'll split the data into training and test sets to evaluate the model's performance on unseen data.
\end{enumerate}

Let's start with encoding the categorical variables and scaling the numerical variables. This preparation will finalize the dataset for machine learning. \\

\generatedCode
\begin{lstlisting}
from sklearn.model_selection import train_test_split
from sklearn.preprocessing import StandardScaler, OneHotEncoder
import pandas as pd

# Load the prepared data
data = pd.read_csv("kidney_prepared.csv")

# Separate features and target variable
X = data.drop('eGFR12', axis=1)
y = data['eGFR12']

# Encoding categorical variables
categorical_cols = ['Donor_First_A_Broad', 'Donor_Second_A_Broad', 'Donor_First_B_Broad', 'Donor_Second_B_Broad', 'Donor_First_C_Broad', 'Donor_First_DR_Broad', 'Donor_Second_DR_Broad', 'Donor_First_DQ_Broad', 'Donor_Second_DQ_Broad', 'Recip_First_A_Broad', 'Recip_Second_A_Broad', 'Recip_First_B_Broad', 'Recip_Second_B_Broad', 'Recip_First_DR_Broad', 'Recip_Second_DR_Broad', 'Recip_First_DQ_Broad', 'Recip_Second_DQ_Broad']
encoder = OneHotEncoder(sparse=False, handle_unknown='ignore')
X_encoded = pd.DataFrame(encoder.fit_transform(X[categorical_cols]))
X_encoded.columns = encoder.get_feature_names(categorical_cols)
X = X.drop(categorical_cols, axis=1).reset_index(drop=True)
X = pd.concat([X, X_encoded], axis=1)

# Scaling numerical variables
scaler = StandardScaler()
numerical_cols = X.select_dtypes(include=['int64', 'float64']).columns
X[numerical_cols] = scaler.fit_transform(X[numerical_cols])

# Splitting the dataset into training and test sets
X_train, X_test, y_train, y_test = train_test_split(X, y, test_size=0.2, random_state=42)

# Save the processed data
X_train.to_csv('X_train.csv', index=False)
X_test.to_csv('X_test.csv', index=False)
y_train.to_csv('y_train.csv', index=False)
y_test.to_csv('y_test.csv', index=False)

print("Categorical variables encoded, numerical variables scaled, and data split into training and test sets.")
\end{lstlisting}
\end{assistantbox}

\mydots

\noindent \textbf{Summary of work after this point:} The system then proceeds to work with the user on model building and utilization, without returning to feature selection.


\newpage
\subsubsection*{Example 3: failure to check for data leakage}

\noindent In this example, we observe that \baseline~tends not to pick up on the presence of likely data leakage (or, indeed, the identifier) columns, as it lacks a structured set of checks in the workflow that would ensure this.

\mygap

\noindent \textbf{Summary of work up to this point:} The user's data was uploaded, an exploratory data analysis performed, and missing values were handled. No discussion of the features that should be removed from the outset was initiated by the system, so the set of features being used at this point is large, at 274.

\mydots

\begin{assistantbox}
\assistant The dataset now has 274 columns after encoding categorical variables and imputing missing values for both continuous and categorical variables. This transformation ensures that the dataset is ready for regression analysis, with all variables in a format suitable for machine learning models. \\

The next steps in our analysis will be:

\begin{enumerate}
    \item \textbf{Feature Selection}: We'll identify the most relevant features for predicting \texttt{eGFR12}. This can be done through various methods such as correlation analysis, or more sophisticated techniques like feature importance from tree-based models. Given the expanded feature set post-encoding, this step is crucial to simplify the model and potentially improve its performance.
    \item \textbf{Model Training}: We'll select a regression model to train on our dataset. Common choices include linear regression for its simplicity and interpretability, or tree-based models like Random Forest or Gradient Boosting for potentially better performance at the cost of some interpretability.
    \item \textbf{Model Evaluation}: We'll evaluate the model using appropriate metrics. For regression tasks, common metrics include Mean Absolute Error (MAE), Mean Squared Error (MSE), and R-squared (R$^2$).
\end{enumerate}

Would you like to proceed with feature selection next, or do you have any specific preferences for the model training step?
\end{assistantbox}

\begin{userbox}
\user Yes
\end{userbox}

\begin{assistantbox}
\assistant Great! Let's proceed with feature selection to identify the most relevant features for predicting \texttt{eGFR12}. Given the large number of features created after one-hot encoding, feature selection will help us reduce the dimensionality of the dataset and potentially improve the model's performance by focusing on the most informative variables. \\

A straightforward and effective approach for feature selection in regression tasks is to use a tree-based model, like a Random Forest, to assess feature importance. Random Forests are useful for this purpose because they provide a measure of how much each feature contributes to the prediction accuracy, without requiring the data to be linearly separable or normally distributed. \\

Here's the plan:

\begin{enumerate}
    \item Train a basic Random Forest regressor on the dataset.
    \item Use the feature importances provided by the model to select a subset of the most important features.
\end{enumerate}

Let's start by training a Random Forest regressor and obtaining the feature importances. \\

\generatedCode
\vspace{0.3cm}
\\
\textit{Dependencies:}
\begin{lstlisting}
pandas
sklearn
\end{lstlisting}

\begin{lstlisting}
import pandas as pd
from sklearn.ensemble import RandomForestRegressor
from sklearn.model_selection import train_test_split

# Load the encoded dataset
data = pd.read_csv("preprocessed_data_encoded.csv")

# Separate the features and target variable
X = data.drop("eGFR12", axis=1)
y = data["eGFR12"]

# Split the data into training and testing sets
X_train, X_test, y_train, y_test = train_test_split(X, y, test_size=0.2, random_state=42)

# Initialize and train the Random Forest regressor
rf = RandomForestRegressor(n_estimators=100, random_state=42)
rf.fit(X_train, y_train)

# Get feature importances
feature_importances = rf.feature_importances_

# Create a DataFrame to view the feature importances
features_df = pd.DataFrame({'Feature': X.columns, 'Importance': feature_importances})
features_df = features_df.sort_values(by='Importance', ascending=False)

# Save the feature importances to a CSV for review
features_df.to_csv("feature_importances.csv", index=False)

print("Feature importances saved to feature_importances.csv")
\end{lstlisting}

\end{assistantbox}

\begin{execbox}
\codeExecOut
\codeExecFail
\begin{lstlisting}[style=codeoutput]
error: subprocess-exited-with-error
[...]
    The 'sklearn' PyPI package is deprecated, use 'scikit-learn' rather than 'sklearn' for pip commands.
[...]
note: This is an issue with the package mentioned above, not pip.
\end{lstlisting}
\end{execbox}

\noindent The system encountered an error in dependency specification but recovered by correcting the dependency name from \texttt{sklearn} to \texttt{scikit-learn}.

\begin{execbox}
\codeExecOut
\codeExecSuccess
\begin{lstlisting}[style=codeoutput]
Feature importances saved to feature_importances.csv
\end{lstlisting}
\end{execbox}

\begin{assistantbox}
\assistant The feature importances have been successfully calculated and saved to \texttt{feature\_importances.csv}. This file contains a list of features ranked by their importance in predicting \texttt{eGFR12}, as determined by the Random Forest regressor. \\

With this information, we can:

\begin{itemize}
    \item Select a subset of the most important features for the regression model to simplify the model and potentially improve performance.
    \item Proceed to model training with the selected features.
\end{itemize}

Would you like to review the feature importances and select a specific number of top features for the model, or should we proceed with all the features available?
\end{assistantbox}

\begin{userbox}
\user Yes, review them
\end{userbox}

\begin{assistantbox}
\assistant To review the feature importances, I'll generate code to load and display the top features from the \texttt{feature\_importances.csv} file. This will help us decide how many features we might want to include in our regression model based on their importance scores. \\

\generatedCode
\begin{lstlisting}
import pandas as pd

# Load the feature importances
feature_importances = pd.read_csv("feature_importances.csv")

# Display the top 20 features by importance
print("Top 20 features by importance:")
print(feature_importances.head(20))
\end{lstlisting}

\end{assistantbox}

\begin{execbox}
\codeExecOut
\codeExecSuccess
\begin{lstlisting}[style=codeoutput]
Top 20 features by importance:
                    Feature  Importance
0                     eGFR3    0.506512
1                      RAGE    0.170737
2                     GSURV    0.027731
3                      DAGE    0.024606
4                      DBMI    0.020644
5                  CIT_MINS    0.015304
6                  RECIP_ID    0.013962
7                 WAIT_TIME    0.012750
8                DRET_CREAT    0.011237
9                     TX_ID    0.008558
10                     DCOD    0.008022
11                      PRD    0.006450
12                     DSEX    0.005084
13                  RETHNIC    0.004640
14  Donor_First_C_Broad_Cw7    0.004598
15                 DONOR_ID    0.004161
16                   CRF_TX    0.002980
17  Recip_First_B_Broad_B40    0.002926
18              organs_dntd    0.002925
19               ORGANS_TXD    0.002867
\end{lstlisting}
\end{execbox}

\noindent Notice that the columns eGFR3 and GSURV appear as important features, which is unsurprising, since these represent a data leakage for the target eGFR12. The identifier columns such as RECIP\_ID are also present.

\noindent The system then continues through to the model building stage with the user, using the top 20 features identified here.

\mydots

\noindent After the model building stage, the user requested the feature importance values of the final trained model.

\begin{userbox}
\user I would like to see feature importances
\end{userbox}

\begin{assistantbox}
\assistant To provide insights into which features are most important for predicting \texttt{eGFR12} according to the optimized Random Forest model, let's extract and display the feature importances from the trained model. \\

We'll generate code to load the optimized model, extract the feature importances, and then display them in descending order of importance. \\

\generatedCode
\begin{lstlisting}
import pandas as pd
import joblib

# Load the optimized Random Forest model
rf_optimized = joblib.load('rf_optimized_model.joblib')

# Load the dataset to get the feature names
data = pd.read_csv("data_selected_features.csv")
feature_names = data.drop('eGFR12', axis=1).columns

# Extract feature importances from the model
feature_importances = rf_optimized.feature_importances_

# Create a DataFrame to display them
importances_df = pd.DataFrame({'Feature': feature_names, 'Importance': feature_importances})
importances_df = importances_df.sort_values(by='Importance', ascending=False)

# Display the feature importances
print(importances_df)
\end{lstlisting}

\end{assistantbox}

\begin{execbox}
\codeExecOut
\codeExecSuccess
\begin{lstlisting}[style=codeoutput]
Feature  Importance
0                     eGFR3    0.370966
1                      RAGE    0.182343
3                      DAGE    0.114246
12                     DSEX    0.047448
2                     GSURV    0.047023
4                      DBMI    0.038885
7                 WAIT_TIME    0.032089
9                     TX_ID    0.030701
8                DRET_CREAT    0.030360
6                  RECIP_ID    0.030002
5                  CIT_MINS    0.027119
10                     DCOD    0.022366
11                      PRD    0.017126
13                  RETHNIC    0.005409
14  Donor_First_C_Broad_Cw7    0.003918
\end{lstlisting}
\end{execbox}

\noindent Once again, we notice that in the final trained model, the data leakage columns (eGFR3, GSURV) persist, making the model not practically applicable for new predictions, as these variables are not going to be available for patients at transplant time. Since they were not removed, some ID columns (RECIP\_ID, TX\_ID) are also present, which may represent the model overfitting on the particular set of IDs used in the training data, and likely misleading feature importance values.

\mydots

\noindent \textbf{Summary of work after this point:} The system then proceeded to work with the user on the final task, that is, subgroup analysis by the DSEX variable.


\newpage
\subsubsection{\textit{CliMB} examples}

An example of an entire \textit{CliMB} session can be watched here: \url{https://youtu.be/76XuR0K3F5Y}. We provide examples below that show the advantages of the advanced reasoning approach and other features of \textit{CliMB}. The below examples are sourced from a different session from the recording and hence there are minor differences.

\subsubsection*{Example 1: Data exploration}

\noindent This example demonstrates \textit{CliMB}'s approach to getting a detailed and relevant data analysis via feedback from the user and the available tools (EDA and descriptive statistics).

\mygap

\noindent \textbf{Summary of work up to this point:} The user's file has been loaded, and the alignment of the system's capabilities and the user's intended research question has been confirmed.

\mydots

\begin{assistantbox}
\assistant \texttt{Planning step}
\end{assistantbox}

\begin{assistantbox}
\assistant To begin our exploratory data analysis, let's first identify all the columns in your dataset. This will allow us to understand the structure of your data better and decide if there are any columns you would like to exclude from the analysis or if you prefer to only keep certain columns. \\

\generatedCode
\begin{lstlisting}
import pandas as pd

# Load the dataset
data = pd.read_csv("kidney_train.csv")

# Print the names of all the columns, one per line for clarity
print("List of all columns in the dataset:")
for column in data.columns:
    print(column)
\end{lstlisting}
\end{assistantbox}

\begin{execbox}
\codeExecOut
\codeExecSuccess
\begin{lstlisting}[style=codeoutput]
List of all columns in the dataset:
DONOR_ID
DAGE
DSEX

... shortened for brevity ...

Recip_Second_DQ_Broad
Recip_Second_DQ_Split
recip_first_Bw
\end{lstlisting}
\end{execbox}

\begin{assistantbox}
\assistant Given the list of columns in your dataset, would you like to exclude certain columns from the analysis, or would you prefer to only keep certain columns? If so, could you specify which columns you would like to exclude or keep? This will help tailor the dataset to your specific analysis needs.
\end{assistantbox}

\noindent The system provides an opportunity for the user to review the features in the dataset early in the pipeline, and to select a subset of these as needed. In this example, the clinician scientist required a fairly complex preprocessing step in terms of which columns were to be removed; this was requested and accomplished by the system.

\begin{userbox}
\user I want to remove some unnecessary columns. These are: all columns that have a name like $<$letter$>$\_$<$number$>$, all columns that start with Recip\_ or Donor\_, and also remove columns ``ALI\_0'', ``OKI\_0'', ``recip\_first\_Bw''.
\end{userbox}

\begin{assistantbox}
\assistant I'll now generate the code to exclude the specified columns from your dataset. This includes removing all columns with names matching the pattern \texttt{$<$letter$>$\_$<$number$>$}, all columns starting with \texttt{Recip\_} or \texttt{Donor\_}, and specifically the columns \texttt{ALI\_0}, \texttt{OKI\_0}, and \texttt{recip\_first\_Bw}. The modified dataset will be saved with a suffix \texttt{\_user\_cols} in the filename. \\

\generatedCode
\begin{lstlisting}
import pandas as pd
import re

# Load the dataset
data = pd.read_csv("kidney_train.csv")

# Define the patterns for columns to be removed
patterns = [r'^[A-Z]_\d$', r'^Recip_', r'^Donor_', r'^ALI_0$', r'^OKI_0$', r'^recip_first_Bw$']

# Compile the patterns into regex objects
regexes = [re.compile(pattern) for pattern in patterns]

# Identify columns to remove based on the patterns
columns_to_remove = [column for column in data.columns if any(regex.match(column) for regex in regexes)]

# Remove the identified columns
data_modified = data.drop(columns=columns_to_remove)

# Save the modified dataset
data_modified.to_csv("kidney_train_user_cols.csv", index=False)

print(f"Modified dataset saved as 'kidney_train_user_cols.csv'. Removed {len(columns_to_remove)} columns.")
\end{lstlisting}
\end{assistantbox}

\begin{execbox}
\codeExecOut
\codeExecSuccess
\begin{lstlisting}[style=codeoutput]
Modified dataset saved as 'kidney_train_user_cols.csv'. Removed 52 columns.
\end{lstlisting}
\end{execbox}

\begin{toolbox}
\toolExecOut{EDA} \\

\toolLogs
\begin{lstlisting}[style=codeoutput]
Getting dataset basic info...
Getting descriptive statistics for numerical features...
Getting detailed information on categorical variables...
Performing missing values analysis...
Identified numeric columns that should be considered categorical:
['DSEX', 'DETHNIC', 'HYPERTENSION', 'DGRP', 'RSEX', 'RETHNIC', 'GRAFT_NO', 'TX_YR', 'HLA_MM', 'organ_type', 'REJECTION', 'DGF', 'GCENS', 'COF', 'TX_TYPE', 'organs_dntd', 'ORGANS_TXD', 'PAST_DIABETES', 'PAST_HYPERTENSION']
Performing correlation analysis...
Performing potential outliers identification...
Performing duplicate records analysis...
\end{lstlisting}

\toolOut
\begin{lstlisting}[style=codeoutput]
Dataset Shape: 1200 rows and 36 columns
Column Names and Types:
DONOR_ID              object
DAGE                   int64
DSEX                   int64
... shortened for brevity ...
ORGANS_TXD             int64
PAST_DIABETES        float64
PAST_HYPERTENSION    float64

Descriptive Statistics for Numerical Features:
[shortened here for brevity]
            DONOR_ID         DAGE         DSEX     ...   ORGANS_TXD  PAST_DIABETES  PAST_HYPERTENSION
count    1200.000000  1200.000000  1200.000000     ...  1200.000000    1162.000000        1153.000000
mean    64055.581667    44.575833     1.464167     ...     3.371667       1.036145           1.193408
std     13818.714125    16.002531     0.498922     ...     1.168904       0.186730           0.395142
[...]

Identified numeric value columns that should most likely be considered categoricals:
['DSEX', 'DETHNIC', 'HYPERTENSION', 'DGRP', 'RSEX', 'RETHNIC', 'GRAFT_NO', 'TX_YR', 'HLA_MM', 'organ_type', 'REJECTION', 'DGF', 'GCENS', 'COF', 'TX_TYPE', 'organs_dntd', 'ORGANS_TXD', 'PAST_DIABETES', 'PAST_HYPERTENSION'].
This is done by checking whether the column contains only integers and has a low number of unique values (<20 or <5% of total examples).

Detailed Information on Categorical Variables:
GRAFT_NO - Unique Values: 4 
Top 5 Values:
GRAFT_NO
1    870
2    279
3     48
4      3

TX_TYPE - Unique Values: 1 
Top 5 Values:
TX_TYPE
10    1200

RECIP_ID - Unique Values: 1200 
Top 5 Values:
RECIP_ID
ZGZWPAG    1
QUUFRYX    1
95WCXRJ    1
NJ96Z1C    1
TYVD858    1
Rare Categories:
RECIP_ID
9IV3SFC    1
I1PXOJP    1
QWJIG75    1
A0IBUOM    1
EMHWBRM    1

... shortened for brevity ...

organ_type - Unique Values: 2 
Top 5 Values:
organ_type
12    608
11    592

DGF - Unique Values: 3 
Top 5 Values:
DGF
1.0    546
2.0    230
3.0     27

PAST_DIABETES - Unique Values: 2 
Top 5 Values:
PAST_DIABETES
1.0    1135
2.0      28

Missing Values Analysis:
DETHNIC                 6
HYPERTENSION           79
DRET_CREAT            109
DBMI                   59
RETHNIC                16
RBMI                  855
HLA_MM               1200
CIT_MINS               18
eGFR3                 217
eGFR12                197
REJECTION              72
DGF                   397
GSURV                   2
GCENS                   2
COF                   774
RCOD                 1017
PAST_DIABETES          37
PAST_HYPERTENSION      49

Count of columns with all NaN values: 1
Correlation Analysis:

Most Positively Correlated Features:
       Feature 1          Feature 2  Correlation
0    organs_dntd         ORGANS_TXD     0.910547
1          eGFR3             eGFR12     0.879036
2       GRAFT_NO             CRF_TX     0.658768
3      WAIT_TIME              TX_YR     0.446634
4          TX_YR             CRF_TX     0.405210
5       GRAFT_NO              TX_YR     0.395091
6           DAGE               RAGE     0.375703
7          TX_YR                DGF     0.362208
8      WAIT_TIME             CRF_TX     0.358320
9  PAST_DIABETES  PAST_HYPERTENSION     0.325089

Most Negatively Correlated Features:
      Feature 1 Feature 2  Correlation
0         TX_YR     GSURV    -0.486558
1          DAGE    eGFR12    -0.460534
2          DAGE     eGFR3    -0.435481
3  HYPERTENSION      DGRP    -0.427371
4         GSURV     GCENS    -0.426957
5          RBMI      RCOD    -0.392979
6          DSEX    eGFR12    -0.369204
7          DSEX     eGFR3    -0.354643
8          RAGE    eGFR12    -0.350899
9          RAGE     eGFR3    -0.345405

Outlier Identification for Numerical Features:
DAGE - Outliers Count: 0
[Lower Bound: -4.5, Upper Bound: 94.5]
DSEX - Outliers Count: 0
[Lower Bound: -1.75, Upper Bound: 2.75]
DETHNIC - Outliers Count: 41
[Lower Bound: 0, Upper Bound: 0]
... shortened for brevity ...
ORGANS_TXD - Outliers Count: 121
[Lower Bound: 0.25, Upper Bound: 4.75]
PAST_DIABETES - Outliers Count: 65
[Lower Bound: 0, Upper Bound: 0]
PAST_HYPERTENSION - Outliers Count: 275
[Lower Bound: 0, Upper Bound: 0]

Duplicate Records: 0
\end{lstlisting}

\vspace{1cm}
\toolReport \\

Here is a correlogram showing the correlation between features: \\

\includegraphics[width=0.99\linewidth]{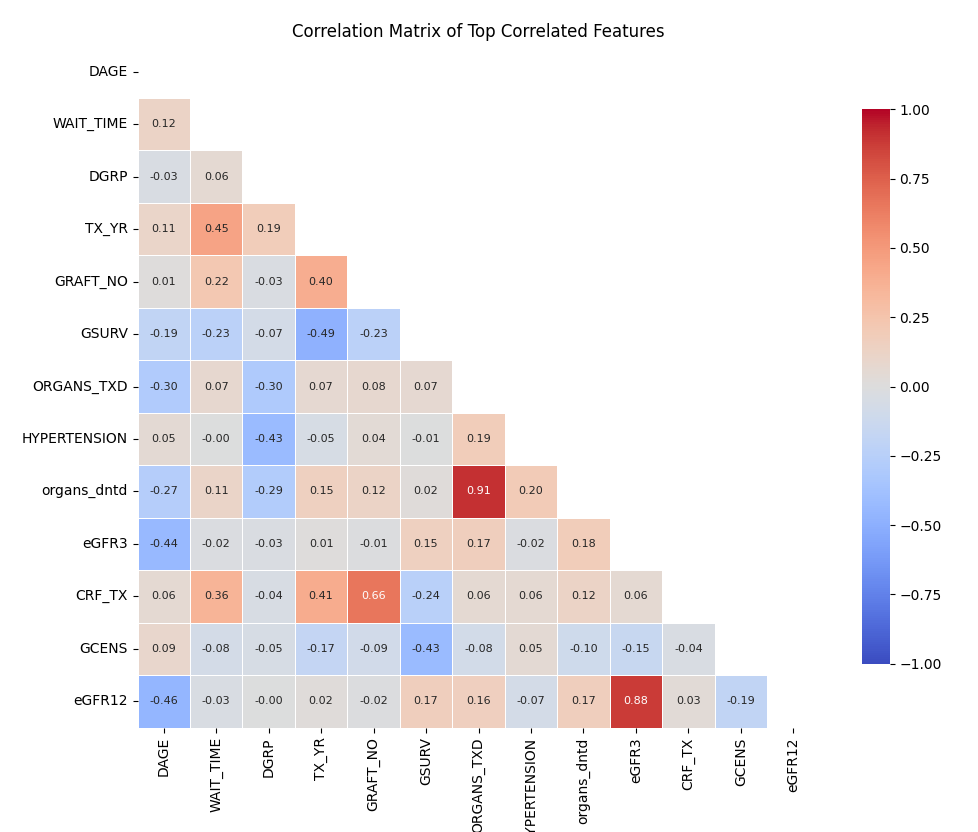} \\

A correlogram is a visual tool that shows the relationships between different variables (or features) in a dataset. It presents a grid of color-coded squares, where each square represents the strength and direction of the relationship between two variables. Brighter or darker colors indicate stronger relationships. Positive relationships (where variables increase together) and negative relationships (where one variable increases as the other decreases) are shown with different colors. Here we use reds to denote positive, and blue to denote negative relationships. This makes it easy to see which pairs of variables are related, and how closely they are connected.

\end{toolbox}

\noindent Notice that \textit{CliMB} tools are able to produce a report that may include text and figures, in order to enhance the user's understanding of the results of the tool execution.

\begin{assistantbox}
\assistant The exploratory data analysis (EDA) has provided us with a wealth of information about your dataset. Here are some key insights: \\

\begin{enumerate}
    \item \textbf{Dataset Overview:} Your dataset consists of 1200 rows and 36 columns after some preprocessing. It includes a mix of numerical and categorical variables.
    \item \textbf{Numerical Feature Analysis:}
        \begin{itemize}
            \item The descriptive statistics for numerical features suggest varying ranges and distributions, which is typical for clinical data. For example, \texttt{DAGE} (donor age) and \texttt{RAGE} (recipient age) show a broad range of values, indicating a diverse sample of donors and recipients.
            \item Some variables like \texttt{eGFR3} and \texttt{eGFR12} (estimated Glomerular Filtration Rate at 3 and 12 months) are crucial for your analysis as they directly relate to the kidney function post-transplant.
        \end{itemize}
    \item \textbf{Categorical Variable Analysis:}
        \begin{itemize}
            \item Several categorical variables have been identified, such as \texttt{DSEX} (donor sex), \texttt{RETHNIC} (recipient ethnicity), and \texttt{TX\_TYPE}. These variables could play significant roles in your predictive model.
            \item The dataset includes many unique identifiers (\texttt{DONOR\_ID}, \texttt{RECIP\_ID}, \texttt{TX\_ID}), which are not useful for modeling but essential for data management.
        \end{itemize}
    \item \textbf{Missing Values Analysis:}
        \begin{itemize}
            \item Certain columns have a significant number of missing values, such as \texttt{RBMI} (recipient BMI) with 855 missing values and \texttt{RCOD} (reason for coding) with 1017 missing values. This indicates a need for careful handling of missing data.
        \end{itemize}
    \item \textbf{Correlation Analysis:}
        \begin{itemize}
            \item The most positively correlated features include variables like \texttt{organs\_dntd} and \texttt{ORGANS\_TXD}, which is expected. Interestingly, \texttt{eGFR3} and \texttt{eGFR12} are highly correlated, suggesting that early graft function is predictive of the 12-month outcome.
        \end{itemize}
    \item \textbf{Outlier Identification:}
        \begin{itemize}
            \item Several features have outliers, which could be genuine variations or errors. For instance, \texttt{DRET\_CREAT} (donor creatinine) and \texttt{CIT\_MINS} (cold ischemia time in minutes) show outliers that might need further investigation.
        \end{itemize}
    \item \textbf{Duplicate Records Analysis:}
        \begin{itemize}
            \item No duplicate records were found, indicating good data integrity.
        \end{itemize}
\end{enumerate}

This analysis has provided a foundational understanding of your dataset's structure, variables, and potential challenges like missing values and outliers. It's a crucial step towards building a robust predictive model. \\

Would you like to proceed with generating descriptive statistics for a more detailed analysis of each variable, including visualizations?
\end{assistantbox}

\begin{userbox}
\user Yes, let's see descriptive statistics.
\end{userbox}

\begin{toolbox}
\toolExecOut{descriptive\_statistics} \\

\toolLogs
\begin{lstlisting}[style=codeoutput]
Creating the descriptive statistics table...
Saving the summary table to:
[...]/session_2024-09-17_18-58-36/kidney_train_user_cols.csv__descriptive_stats.csv
Identified numeric columns that should be considered categorical:
['DSEX', 'DETHNIC', 'HYPERTENSION', 'DGRP', 'RSEX', 'RETHNIC', 'GRAFT_NO', 'TX_YR', 'HLA_MM', 'organ_type', 'REJECTION', 'DGF', 'GCENS', 'COF', 'TX_TYPE', 'organs_dntd', 'ORGANS_TXD', 'PAST_DIABETES', 'PAST_HYPERTENSION']
Creating plots for the data...
Normally distributed features:
['RBMI']
Not normally distributed features:
['CIT_MINS', 'CRF_TX', 'DAGE', 'DBMI', 'DCOD', 'DRET_CREAT', 'GSURV', 'PRD', 'RAGE', 'RCOD', 'WAIT_TIME', 'eGFR12', 'eGFR3']
Plotted a bar plot for: 'COF'
Plotted a bar plot for: 'DETHNIC'
Plotted a bar plot for: 'DGF'
... shortened for brevity ...
Plotted a histogram and box plot for: 'WAIT_TIME'
Plotted a histogram and box plot for: 'eGFR12'
Plotted a histogram and box plot for: 'eGFR3'
\end{lstlisting}

\toolOut
\begin{lstlisting}[style=codeoutput]
Variable             Summary
DONOR_ID                             
    WEE6K0G              1/1200 (0.1)
    5LC7OU5              1/1200 (0.1)
    ZCGNZ7F              1/1200 (0.1)
    QHOA7GH              1/1200 (0.1)
    FRPID0N              1/1200 (0.1)
    Other            1195/1200 (99.6)
DAGE               47.0 (34.0 - 56.0)
DSEX                                 
    1                 632/1200 (52.7)
    2                 568/1200 (47.3)
... shortened for brevity ...
ORGANS_TXD                           
    3                 563/1200 (46.9)
    2                 210/1200 (17.5)
    4                 208/1200 (17.3)
    5                   98/1200 (8.2)
    6                   84/1200 (7.0)
    Other               37/1200 (3.1)
PAST_DIABETES                        
    1.0              1135/1163 (97.6)
    2.0                 28/1163 (2.4)
PAST_HYPERTENSION                    
    1.0               925/1151 (80.4)
    2.0               226/1151 (19.6)

The following plots have also been created and saved:
- COF: descr__bar_plot__COF.png
- DETHNIC: descr__bar_plot__DETHNIC.png
- DGF: descr__bar_plot__DGF.png
... shortened for brevity ...
- WAIT_TIME: descr__hist_box_plot__WAIT_TIME.png
- eGFR12: descr__hist_box_plot__eGFR12.png
- eGFR3: descr__hist_box_plot__eGFR3.png
\end{lstlisting}

\toolReport \\

To view the plots, please select any of the images in the "Working Directory" tab on the right.
\end{toolbox}

\begin{assistantbox}
\assistant The descriptive statistics have provided a detailed overview of each variable in your dataset. Here's a summary of the key findings:

\begin{enumerate}
    \item \textbf{Variable Summaries:}
        \begin{itemize}
            \item Categorical variables such as \texttt{DSEX} (donor sex), \texttt{RETHNIC} (recipient ethnicity), and \texttt{TX\_TYPE} (transplant type) show a clear distribution of values, which could be important for your analysis.
            \item Numerical variables such as \texttt{DAGE} (donor age), \texttt{RAGE} (recipient age), and \texttt{eGFR12} (estimated Glomerular Filtration Rate at 12 months) have been summarized with median and interquartile ranges, providing insights into their distribution.
        \end{itemize}
    \item \textbf{Outliers and Missing Values:}
        \begin{itemize}
            \item The analysis has identified potential outliers and missing values in several variables. This is crucial for data cleaning and preprocessing to ensure the quality of your predictive model.
        \end{itemize}
    \item \textbf{Plots Generated:}
        \begin{itemize}
            \item Various plots have been generated to visually summarize the data, including bar plots for categorical variables and histograms and box plots for numerical variables. These visualizations can help identify patterns, outliers, and distributions in your dataset.
        \end{itemize}
\end{enumerate}

Here are five important plots that you might find useful:\\

\begin{itemize}
    \item Estimated Glomerular Filtration Rate at 12 months (\texttt{eGFR12}): \\
\includegraphics[width=1.0\linewidth]{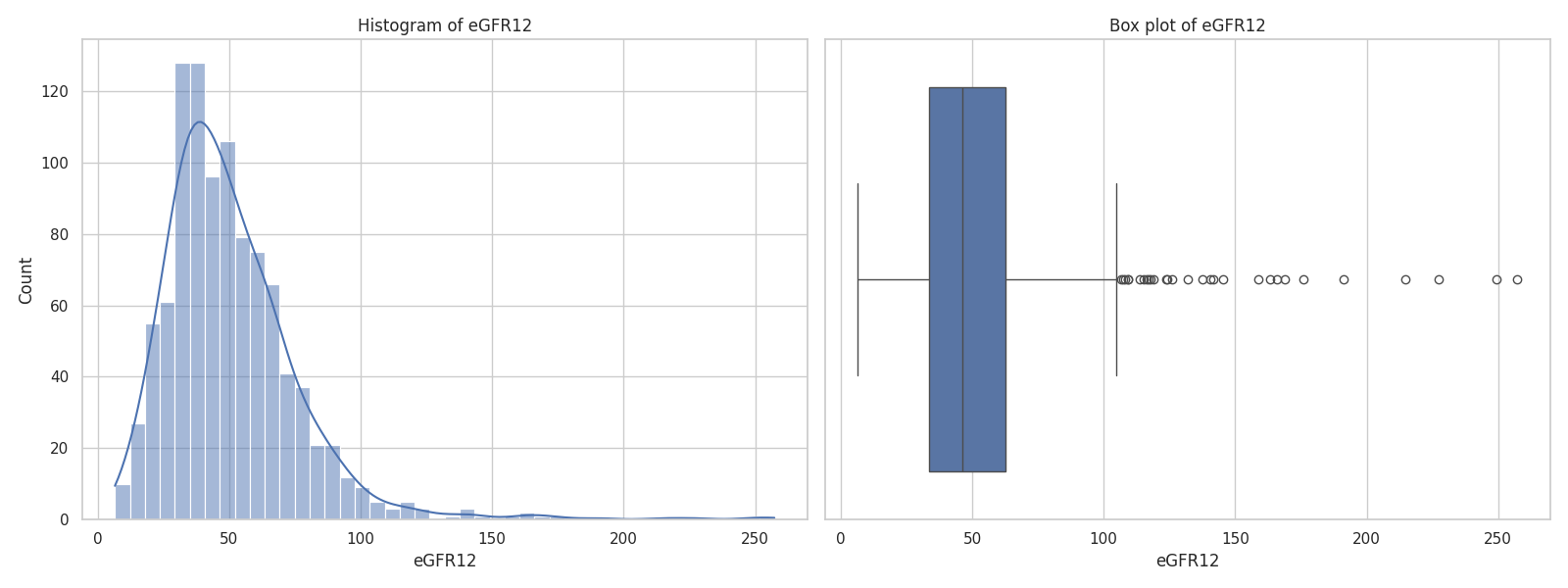}
    \item Donor age (\texttt{DAGE}): \\
\includegraphics[width=1.0\linewidth]{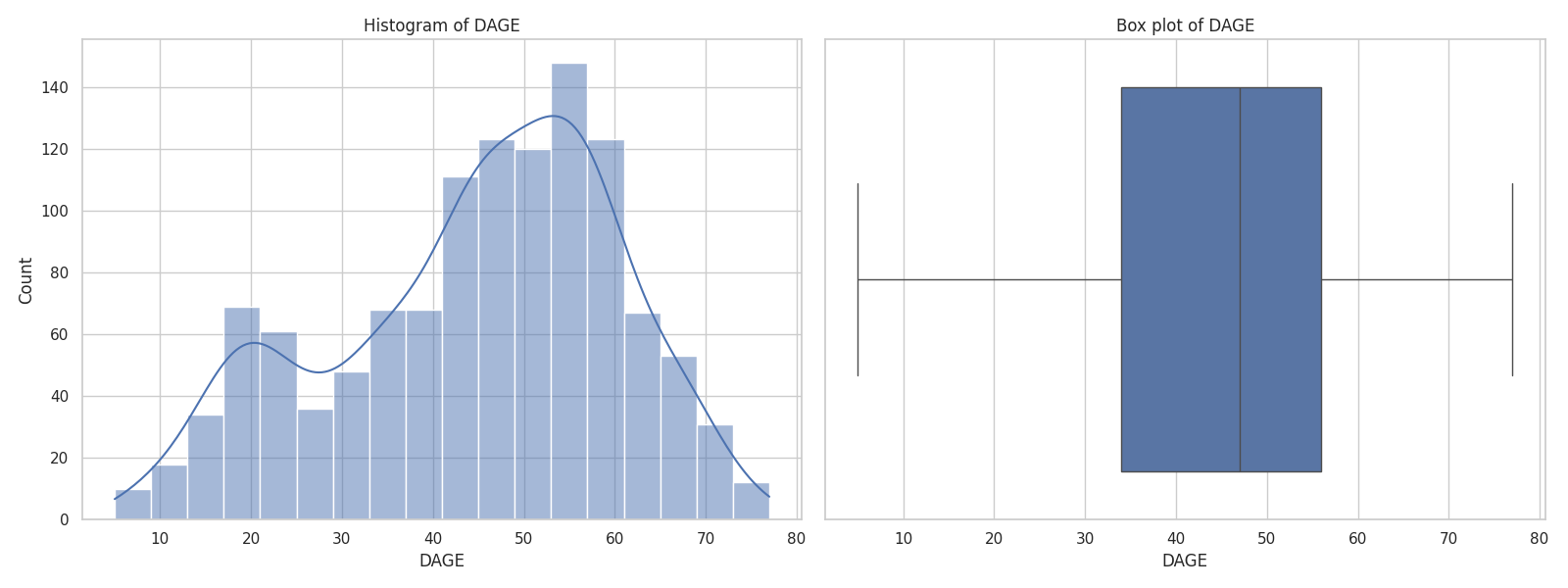}
    \item Recipient age (\texttt{RAGE}): \\
\includegraphics[width=1.0\linewidth]{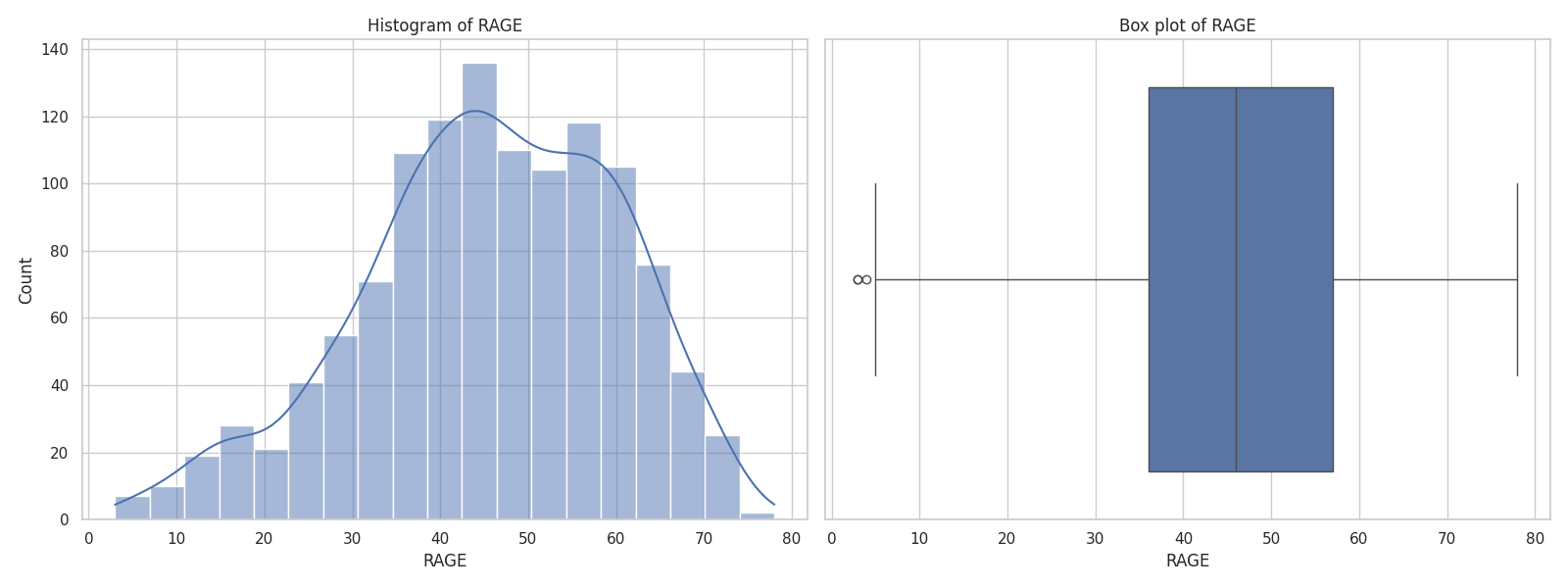}
    \item Cold Ischemia Time in minutes (\texttt{CIT\_MINS}): \\
\includegraphics[width=1.0\linewidth]{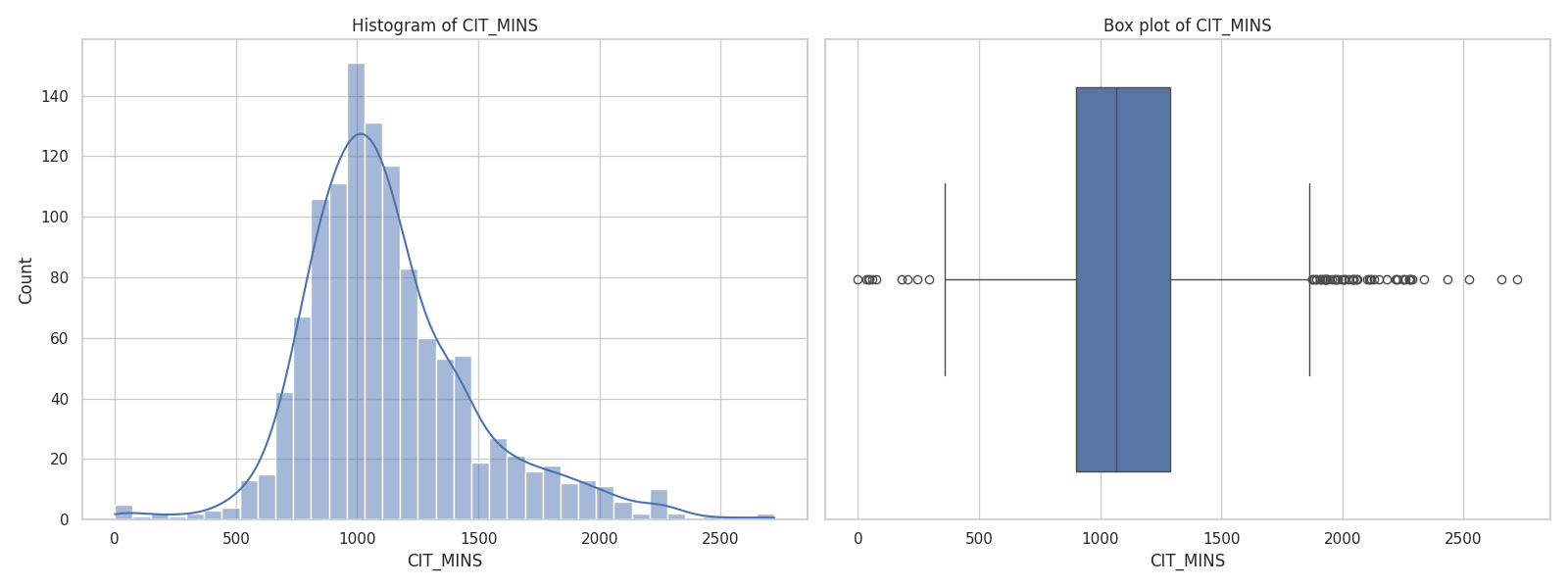}
    \item Cause of Failure (\texttt{COF}): \\
\includegraphics[width=1.0\linewidth]{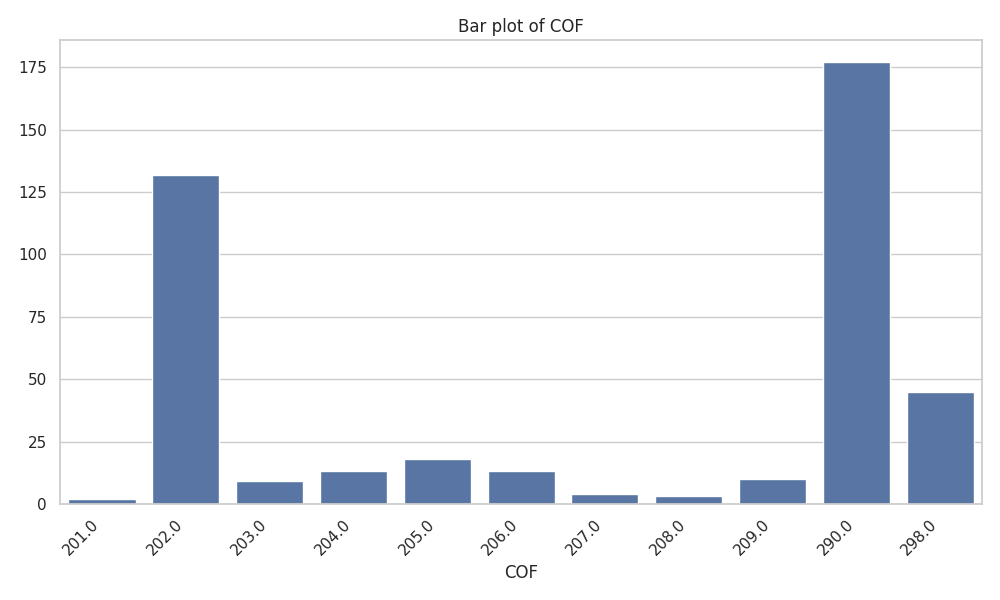}
\end{itemize}

These plots provide a visual representation of the key variables in your dataset, offering insights into their distribution and potential impact on graft function at 12 months post-transplant. \\

Please take a moment to review these plots and the descriptive statistics. If you have any questions or need further analysis, feel free to ask.

\end{assistantbox}

\begin{userbox}
\user That makes sense. Let's continue.
\end{userbox}

\begin{assistantbox}
\assistant \cmark~Task completed
\end{assistantbox}

\mydots

\noindent \textbf{Summary of work after this point:} The system then proceeded to work with the user on the remaining stages of the project: data engineering, model building, and model exploitation.


\subsubsection*{Example 2: Handling of missing values}

\noindent The following example demonstrates how \textit{CliMB} utilizes a multi-step approach to handling missing data, avoiding common missing data pitfalls, and querying user input when necessary using code generation and available tools.

\mygap

\noindent \textbf{Summary of work up to this point:} User data has been loaded, the alignment between the system's capabilities and the user's intended research question has been confirmed, exploratory data analysis and descriptive statistics generation have been completed, and background information about different columns has been discussed with the user.

\noindent At this point, the dataset has 36 columns, 1200 rows, and 5106 (11.82\% of total) missing values.

\mydots

\begin{assistantbox}
\assistant \texttt{Planning step}
\end{assistantbox}

\begin{assistantbox}
\assistant To begin addressing the missing data in your dataset, we'll first need to represent any missing data as \texttt{numpy.nan} values. This standardization is crucial for the subsequent steps of handling missing data effectively. \\

Based on the EDA performed earlier, we identified several columns with missing values, such as \texttt{DETHNIC}, \texttt{HYPERTENSION}, \texttt{DRET\_CREAT}, \texttt{DBMI}, \texttt{RETHNIC}, \texttt{RBMI}, \texttt{HLA\_MM}, \texttt{CIT\_MINS}, \texttt{eGFR3}, \texttt{eGFR12}, \texttt{REJECTION}, \texttt{DGF}, \texttt{GSURV}, \texttt{GCENS}, \texttt{COF}, \texttt{RCOD}, \texttt{PAST\_DIABETES}, and \texttt{PAST\_HYPERTENSION}. It was also noted that there are columns with significant amounts of missing data, which might be represented in non-standard formats. \\

Before proceeding with converting non-standard NaN representations to \texttt{numpy.nan}, I'd like to confirm if there are any specific non-standard values used to represent missing data in these columns (e.g., placeholders like "NA", "unknown", "-1", etc.). Could you confirm if such non-standard missing value indicators are present in the dataset?

\end{assistantbox}

\noindent Note that the workflow allows the system to make reasonable inferences in terms of what placeholders, if any, are used to represent missing values in the dataset; this is done by analysing the relevant project steps completed up to this point (especially exploratory data analysis results). Since in this case, there are in fact no placeholders that represent missing values in the dataset, the system did not suggest any particular placeholders to convert to \texttt{numpy.nan}, however, this was first explicitly checked with the user.

\begin{userbox}
\user I believe we don't have such placeholder values in our dataset.
\end{userbox}

\begin{assistantbox}
\assistant Given that there are no non-standard placeholders for missing values in your dataset, we can proceed directly with the next steps of handling missing data. \\

The next step is to generate code to display per-column percentages of missing values in descending order. This will help us identify columns with a high percentage of missing data, which might need to be dropped or require special attention for imputation. \\

Let's start by displaying the percentage of missing values for each column: \\

\generatedCode
\begin{lstlisting}
import pandas as pd
import numpy as np

# Load the dataset
data = pd.read_csv("kidney_train_user_cols.csv")

# Calculate the percentage of missing values for each column
missing_percentage = data.isnull().mean() * 100

# Display the percentages in descending order
missing_percentage_sorted = missing_percentage.sort_values(ascending=False)
print("Percentage of missing values per column, in descending order:")
print(missing_percentage_sorted)

# Save this information into a CSV file for reference
missing_percentage_sorted.to_csv("missing_percentage.csv", header=True)

# Set pandas options to display all rows and columns
pd.options.display.max_rows = None
pd.options.display.max_columns = None
\end{lstlisting}

\end{assistantbox}

\begin{execbox}
\codeExecOut
\codeExecSuccess
\begin{lstlisting}[style=codeoutput]
Percentage of missing values per column, in descending order:
HLA_MM               100.000000
RCOD                  84.750000
RBMI                  71.250000
COF                   64.500000
DGF                   33.083333
eGFR3                 18.083333
eGFR12                16.416667
DRET_CREAT             9.083333
HYPERTENSION           6.583333
REJECTION              6.000000
DBMI                   4.916667
PAST_HYPERTENSION      4.083333
PAST_DIABETES          3.083333
CIT_MINS               1.500000
RETHNIC                1.333333
DETHNIC                0.500000
GCENS                  0.166667
GSURV                  0.166667
organs_dntd            0.000000
TX_TYPE                0.000000
ORGANS_TXD             0.000000
organ_type             0.000000
DONOR_ID               0.000000
CRF_TX                 0.000000
DAGE                   0.000000
TX_ID                  0.000000
GRAFT_NO               0.000000
WAIT_TIME              0.000000
PRD                    0.000000
RSEX                   0.000000
RAGE                   0.000000
RECIP_ID               0.000000
DCOD                   0.000000
DGRP                   0.000000
DSEX                   0.000000
TX_YR                  0.000000
dtype: float64
\end{lstlisting}
\end{execbox}

\begin{assistantbox}
\assistant Based on the percentage of missing values per column, we have identified that the column \texttt{HLA\_MM} has 100\% missing values, making it a candidate for removal. Additionally, columns like \texttt{RCOD}, \texttt{RBMI}, \texttt{COF}, and \texttt{DGF} have high percentages of missing values, which may impact the dataset's integrity and the predictive model's performance. \\

Considering the rule of thumb that columns with 80\%+ missing values should generally be removed due to the limited information they can provide, we should consider dropping the \texttt{HLA\_MM} and \texttt{RCOD} columns immediately. \\

Before proceeding with the removal of these columns or any others with a high percentage of missing values, I'd like to ask:

\begin{enumerate}
    \item Would you like to set a different threshold for missing values percentage before deciding to remove a column? The default suggestion is 80\%.
    \item Are there any specific columns with high missing values you would like to keep for further analysis or imputation, despite exceeding the missing values threshold?
\end{enumerate}

Please let me know your preferences regarding these questions, and then we can move forward with the appropriate actions.

\end{assistantbox}

\begin{userbox}
\user Let us use the 80\% missing threshold as you suggest.
\end{userbox}

\begin{assistantbox}
\assistant Based on our agreement to use an 80\% missing threshold for deciding whether to remove a column, we will proceed to drop the \texttt{HLA\_MM} column due to 100\% missing values, and the \texttt{RCOD} column due to 84.75\% missing values. This step will help streamline the dataset by removing columns that are unlikely to contribute valuable information to the analysis due to the high proportion of missing data. \\

Let's generate the code to remove these columns and save the modified dataset: \\

\generatedCode
\begin{lstlisting}
import pandas as pd

# Load the dataset
data = pd.read_csv("kidney_train_user_cols.csv")

# Drop columns with more than 80% missing values
columns_to_drop = ['HLA_MM', 'RCOD']
data_modified = data.drop(columns=columns_to_drop)

# Save the modified dataset
data_modified.to_csv("kidney_train_user_cols_nan.csv", index=False)

print(f"Columns {columns_to_drop} were dropped. Modified dataset saved as 'kidney_train_user_cols_nan.csv'.")
\end{lstlisting}
\end{assistantbox}

\begin{execbox}
\codeExecOut
\codeExecSuccess
\begin{lstlisting}[style=codeoutput]
Columns ['HLA_MM', 'RCOD'] were dropped. Modified dataset saved as 'kidney_train_user_cols_nan.csv'.
\end{lstlisting}
\end{execbox}

\noindent The system then proceeds to investigating the missingness with respect to rows, and also applies specific focus on the target column values.

\begin{assistantbox}
\assistant With the columns \texttt{HLA\_MM} and \texttt{RCOD} removed due to high percentages of missing values, we've streamlined the dataset for more effective analysis and modeling. The next step involves considering whether to drop rows with missing values or to use imputation techniques to fill in these gaps. \\

Given the updated dataset, let's generate code to show: \\

\begin{itemize}
    \item The per-column percentage of missing values,
    \item The percentage of total rows that have missing values.
\end{itemize}

This information will help us decide how to proceed with rows containing missing data, especially concerning the target variable \texttt{eGFR12}. It's important to handle missing values in the target variable carefully, as imputing these could introduce bias or inaccuracies. \\

Let's get an overview of the current state of missing data in the dataset: \\

\generatedCode
\begin{lstlisting}
import pandas as pd

# Load the modified dataset
data = pd.read_csv("kidney_train_user_cols_nan.csv")

# Calculate the percentage of missing values for each column
missing_percentage = data.isnull().mean() * 100

# Calculate the percentage of rows with any missing values
rows_with_missing = (data.isnull().any(axis=1).mean()) * 100

# Display the percentages in descending order for columns
print("Percentage of missing values per column, in descending order:")
print(missing_percentage.sort_values(ascending=False))

# Display the percentage of rows with any missing values
print(f"\nPercentage of total rows with any missing values: {rows_with_missing}%")

# Set pandas options to display all results
pd.options.display.max_rows = None
pd.options.display.max_columns = None
\end{lstlisting}
\end{assistantbox}

\begin{execbox}
\codeExecOut
\codeExecSuccess
\begin{lstlisting}[style=codeoutput]
Percentage of missing values per column, in descending order:
RBMI                 71.250000
COF                  64.500000
DGF                  33.083333
eGFR3                18.083333
eGFR12               16.416667
DRET_CREAT            9.083333
HYPERTENSION          6.583333
REJECTION             6.000000
DBMI                  4.916667
PAST_HYPERTENSION     4.083333
PAST_DIABETES         3.083333
CIT_MINS              1.500000
RETHNIC               1.333333
DETHNIC               0.500000
GCENS                 0.166667
GSURV                 0.166667
organ_type            0.000000
TX_TYPE               0.000000
organs_dntd           0.000000
ORGANS_TXD            0.000000
DONOR_ID              0.000000
CRF_TX                0.000000
TX_YR                 0.000000
DAGE                  0.000000
GRAFT_NO              0.000000
WAIT_TIME             0.000000
PRD                   0.000000
RSEX                  0.000000
RAGE                  0.000000
RECIP_ID              0.000000
DCOD                  0.000000
DGRP                  0.000000
DSEX                  0.000000
TX_ID                 0.000000
dtype: float64

Percentage of total rows with any missing values: 97.25%
\end{lstlisting}
\end{execbox}

\begin{assistantbox}
\assistant Given the significant amount of missing data across various columns and that 97.25\% of the rows contain at least one missing value, it's crucial to carefully consider our next steps to preserve as much valuable data as possible. \\

For the target variable \texttt{eGFR12}, which is essential for your predictive model, it's particularly important to handle missing values judiciously. Imputing the target variable is generally not recommended due to the risk of introducing bias. \\

Considering the high percentage of rows with missing values, dropping all such rows could lead to a substantial reduction in dataset size, potentially compromising the study's validity. Therefore, imputation might be a preferable approach for columns other than the target variable, especially if the missing data is not overwhelmingly concentrated in a few rows. \\

Given these considerations:

\begin{itemize}
    \item Are you okay with dropping rows that have missing values in the target variable \texttt{eGFR12}? This step is necessary as imputing the target variable is not recommended.
    \item For other columns with missing data, would you prefer to attempt imputation to preserve as much data as possible? Imputation can be particularly beneficial if missing data is spread across many rows, minimizing the loss of valuable information.
\end{itemize}

Please let me know how you'd like to proceed with these two points, and then we can take the appropriate actions.
\end{assistantbox}

\begin{userbox}
\user I think imputation makes sense for non-eGFR12 columns. \\

I want to work with the eGFR12 column to prepare it. When the eGFR12 value is missing, then:

\begin{enumerate}
    \item If the corresponding GSURV value is between 0 and 364 (inclusive) and the corresponding GCENS is 1 then the eGFR12 value should equal 10. This indicates that the graft was lost priort to the endpoint.
    \item If the corresponding GSURV value is between 0 and 364 (inclusive) and the corresponding GCENS is 0 then rows with missing eGFR12 value should be deleted. This indicates that the patient died with a functioning graft.
    \item Leave the other missing values as they are, those are actually missing.
\end{enumerate}
\end{userbox}

\noindent In this case, the clinician scientist decides to perform a fairly complex, custom handling of missing values in the target column \texttt{eGFR12} -- which either manually imputes the value, discards the row, or leaves the missing value as is for later imputation, depending on logical conditions in other columns. The system successfully completes this request and then proceeds to the final step in the process, automated imputation.

\begin{assistantbox}
\assistant Let's proceed with preparing the \texttt{eGFR12} column according to the conditions you've outlined. We'll handle the missing \texttt{eGFR12} values based on the associated \texttt{GSURV} and \texttt{GCENS} values, with the specific logic you've provided:

\begin{enumerate}
    \item If \texttt{eGFR12} is missing and \texttt{GSURV} is between 0 and 364 (inclusive) and \texttt{GCENS} is 1, then set \texttt{eGFR12} to 10 to indicate graft loss prior to the endpoint.
    \item If \texttt{eGFR12} is missing and \texttt{GSURV} is between 0 and 364 (inclusive) and \texttt{GCENS} is 0, then delete these rows to indicate the patient died with a functioning graft.
    \item Leave other missing \texttt{eGFR12} values as they are, treating them as genuinely missing.
\end{enumerate}

This approach allows us to handle missing data in a way that respects the clinical context of your study. Let's generate the code to implement this logic: \\

\generatedCode
\begin{lstlisting}
import pandas as pd
import numpy as np

# Load the dataset
data = pd.read_csv("kidney_train_user_cols_nan.csv")

# Apply conditions for handling missing eGFR12 values
condition1 = (data['eGFR12'].isnull()) & (data['GSURV'] >= 0) & (data['GSURV'] <= 364) & (data['GCENS'] == 1)
condition2 = (data['eGFR12'].isnull()) & (data['GSURV'] >= 0) & (data['GSURV'] <= 364) & (data['GCENS'] == 0)

# Set eGFR12 to 10 under condition 1
data.loc[condition1, 'eGFR12'] = 10

# Drop rows under condition 2
data = data.drop(data[condition2].index)

# Save the modified dataset
data.to_csv("kidney_train_user_cols_nan_prepared.csv", index=False)

print("eGFR12 column prepared and dataset saved as 'kidney_train_user_cols_nan_prepared.csv'.")
\end{lstlisting}

\end{assistantbox}

\begin{execbox}
\codeExecOut
\codeExecSuccess
\begin{lstlisting}[style=codeoutput]
eGFR12 column prepared and dataset saved as 'kidney_train_user_cols_nan_prepared.csv'.
\end{lstlisting}
\end{execbox}

\begin{assistantbox}
\assistant With the \texttt{eGFR12} column now prepared according to the specified conditions and the dataset saved as 'kidney\_train\_user\_cols\_nan\_prepared.csv', we're ready to move forward with the next steps in handling missing data for the rest of the dataset. \\

Given the considerations and your preference for imputation for non-\texttt{eGFR12} columns, we'll proceed with discussing and preparing for the imputation process. Imputation is a critical step to ensure that our dataset is as complete as possible, which can significantly improve the performance of the predictive model. \\

However, before we can invoke any imputation tool, it's essential to ensure there are no entirely empty columns left in the dataset, as they can cause issues during imputation. We've already addressed columns with a high percentage of missing values, so we should be in a good position to proceed. \\

The next step is to discuss the imputation process itself. Imputation can be done using various methods, ranging from simple approaches like mean or median imputation to more complex ones like model-based imputation. Given the complexity of your dataset and the importance of accurately predicting graft function, a more sophisticated imputation method might be appropriate. \\

\textbf{HyperImpute} is a powerful tool that can handle complex imputation tasks by leveraging multiple imputation models and algorithms to find the best approach for the given data. Before we proceed to use HyperImpute, I'll explain what it does: 

\begin{itemize}
    \item \textbf{HyperImpute} automatically explores a range of imputation models and techniques to fill in missing values in the dataset.
    \item It considers the relationships between variables, the pattern of missingness, and the distribution of data to select the most suitable imputation method.
    \item This process is aimed at minimizing the bias and error introduced by missing data, making the imputed dataset more reliable for subsequent analysis.
\end{itemize}

Would you like to proceed with using HyperImpute for imputing the remaining missing values in your dataset, excluding the \texttt{eGFR12} column which we have already addressed?
\end{assistantbox}

\begin{userbox}
\user Yes, please impute.
\end{userbox}

\begin{toolbox}
\toolExecOut{hyperimpute\_imputation} \\

\toolLogs
\begin{lstlisting}[style=codeoutput]
Imputing all columns
Setting up HyperImpute Imputer...
Imputing data...

Field "model_name" has conflict with protected namespace "model_".

You may be able to resolve this warning by setting `model_config['protected_namespaces'] = ()`.

Missing values before imputation:
DONOR_ID               0
DAGE                   0
DSEX                   0
DETHNIC                6
HYPERTENSION          75
DGRP                   0
DRET_CREAT           106
DCOD                   0
DBMI                  56
RECIP_ID               0
RAGE                   0
RSEX                   0
RETHNIC               16
RBMI                 833
PRD                    0
WAIT_TIME              0
GRAFT_NO               0
TX_ID                  0
TX_YR                  0
CIT_MINS              18
CRF_TX                 0
organ_type             0
eGFR3                199
eGFR12                64
REJECTION             67
DGF                  386
GSURV                  2
GCENS                  2
COF                  759
TX_TYPE                0
organs_dntd            0
ORGANS_TXD             0
PAST_DIABETES         34
PAST_HYPERTENSION     48
dtype: int64

HyperImpute: Iterations completed 1/3
HyperImpute: Iterations completed 2/3
HyperImpute: Iterations completed 3/3
Saving imputation model to: [...]/session_2024-09-17_18-58-36/hyperimpute__kidney_train_user_cols_nan_prepared_imputed.pkl

Missing values after imputation:
DONOR_ID             0
DAGE                 0
DSEX                 0
DETHNIC              0
HYPERTENSION         0
DGRP                 0
DRET_CREAT           0
DCOD                 0
DBMI                 0
RECIP_ID             0
RAGE                 0
RSEX                 0
RETHNIC              0
RBMI                 0
PRD                  0
WAIT_TIME            0
GRAFT_NO             0
TX_ID                0
TX_YR                0
CIT_MINS             0
CRF_TX               0
organ_type           0
eGFR3                0
eGFR12               0
REJECTION            0
DGF                  0
GSURV                0
GCENS                0
COF                  0
TX_TYPE              0
organs_dntd          0
ORGANS_TXD           0
PAST_DIABETES        0
PAST_HYPERTENSION    0
dtype: int64

Saving imputed data...
\end{lstlisting}

\toolOut
\begin{lstlisting}[style=codeoutput]
2671 missing values were imputed. The imputed data has been saved to [...]/session_2024-09-17_18-58-36/kidney_train_user_cols_nan_prepared_imputed.csv
\end{lstlisting}

\end{toolbox}

\noindent At this point, the dataset has 34 columns (the two 80\%+ missing columns having been removed), 1163 rows (the missing target rows having been removed), and no missing values - as one would expect after a successful missing data imputation step.

\mydots

\noindent \textbf{Summary of work after this point:} The system then proceeds to work with the user through the machine learning study, and subsequent post-hoc interpretability including feature importance and subgroup analysis by \texttt{DSEX}.


\newpage
\subsection{Video demonstrations}\label{subsection:school_vids}

The two video demonstrations used in the \textit{Cambridge AI in Medicine Summer School} \cite{noauthor_undated-ip} in order to gather clinician feedback are referenced below.

\begin{itemize}
    \item \textbf{Example 1:} Baseline GPT-4 session. \url{https://youtu.be/KyppniOKvoE},
    \item \textbf{Example 2:} CliMB session. \url{https://youtu.be/9u4lleFPMm8}.
\end{itemize}


\subsection{Consent form}\label{subsection:consent_form}

Before participating in the polls described in section \ref{subsection:Clinician_feedback}, participants were provided with the following information:
 
\noindent \textbf{1. Introduction}

You are invited to participate in a research study that aims to understand how an AI-enabled, interactive chatbot can empower clinicians to build their own predictive models. We are particularly interested in your perceptions of the responses provided by this chatbot. Before you decide to participate, it is important that you understand the purpose of this research and what your participation will involve. Please take the time to read the following information carefully.

\noindent \textbf{2. What is the purpose of this research?}

The primary goal of this research project is to assess the utility of various AI-powered, interactive chatbots designed to assist clinicians like you in building predictive models from clinical datasets. Predictive models can serve multiple purposes. For example, they can help identify patients at high risk for requiring acute care during radiotherapy, enabling clinicians to provide targeted attention. Other models have previously been developed to predict the survival benefits of adjuvant therapy for breast cancer patients.
In this study, the chatbot you will see is designed to assist in building such predictive models. However, our focus is on understanding your perception of how the chatbot communicates and displays information. Specifically, we are interested in how you evaluate different tools across several metrics. Please note that you will not be aware of which tool generated which response. This study does not assess your abilities or intelligence; we are solely interested in your opinions on the chatbot responses.

\noindent \textbf{3. What data will be collected and how?}

During this study, we will collect data related to your preferences. You will be presented with a set of questions that clinicians might ask the chatbot, and you will evaluate the responses generated by two different chatbots. The answers will be anonymized, and you will not know the identity of the chatbot that provided the response. The rating will be done via Zoom’s integrated poll feature. If you are unsure how to respond, please select the option that most aligns with your preference. We will ask general questions about your prior experience with ML and your career stage. We will not collect data on demographic information like age or gender, and no sensitive information will be requested.

\noindent \textbf{4. How will the data be used?}

The data collected will be used to evaluate the effectiveness of different chatbots in assisting clinicians in interactively building predictive models. The aggregated results may be published in academic journals, presented at conferences, and included in technical reports. Individual responses will remain anonymous and will not be identifiable in any published or presented materials.

\noindent \textbf{5. How and for how long will the data be stored?}

All data collected during the study will be securely stored in an encrypted format on secure servers. Data will be retained for a period of five years after the conclusion of the study, as required by our data retention policy, after which it will be securely deleted.

\noindent \textbf{6. Anonymity of Responses}

Your participation in this study will remain anonymous. No personally identifiable information will be associated with your responses in any reports generated from this research. Data will only be presented in aggregate form.

\noindent \textbf{7. Data Sharing with Other Researchers}

Anonymized, aggregated data may be shared with other researchers at a later date. Again, individual responses will not be identifiable.

\noindent \textbf{8. Withdrawal of Consent and Data}

You have the right to withdraw from the study at any time without providing reasons. If you choose to withdraw, all data associated with your participation will be deleted. Your refusal to participate or later withdrawal from the study will not result in any penalty. To withdraw your consent and request the deletion of your data, please contact es583@cam.ac.uk via email.

\noindent \textbf{9. Legal Framework}

Your data will be handled according to the principles and rules set by the General Data Protection Regulation (GDPR).

\noindent \textbf{10. Consent}

Please confirm that you have read and understood the information provided above regarding your participation in this research study. By clicking the box, you confirm that you:
\begin{itemize}
    \item Understand the nature and purpose of the study.
    \item Agree to the collection, use, and storage of your data as described above.
    \item Understand that your participation is entirely voluntary, and that you may withdraw at any time without penalty.
    \item Agree to participate in this study.
\end{itemize}